\documentclass[10pt]{revtex4}
\pdfoutput=1
\usepackage{amssymb}
\usepackage{latexsym}
\usepackage{epsfig}
\begin{document}

\title{The birth of the universe in a new  
G-Theory approach}

\author{Alireza Sepehri $^{1,2}$}
\email{alireza.sepehri@uk.ac.ir} \affiliation{ $^{1}$Faculty of
Physics, Shahid Bahonar University, P.O. Box 76175, Kerman,
Iran.\\$^{2}$ Research Institute for Astronomy and Astrophysics of
Maragha (RIAAM), P.O. Box 55134-441, Maragha, Iran. }

\author{Richard Pincak$^{3,4}$}\email{pincak@saske.sk}
\affiliation{ $^{3}$ Institute of Experimental Physics, Slovak Academy of Sciences,
Watsonova 47,043 53 Kosice, Slovak Republic}
\affiliation{ $^{4}$ Bogoliubov Laboratory of Theoretical Physics, Joint
Institute for Nuclear Research, 141980 Dubna, Moscow region, Russia}

\begin{abstract}
Recently,  Padmanabhan has discussed that the expansion of the cosmic space is due to the difference
between the number of degrees of freedom on the boundary surface
and the number of degrees of freedom in a bulk region. Now, a natural question arises that how these 
degrees of freedom are emerged from nothing? We try to address this issue
in a new theory which is more complete than M-theory and reduces to it with some limitations. In M-theory, there isn't any stable object like stable M3-branes that our universe is formed on it and for this reason can't help us to explain cosmological events. In this research, we  propose a new theory, named G -theory which could be  the mother of M-theory and superstring theory. In G-theory, at the beginning, two types of G0-branes, one with positive energy and one with negative energy are produced from nothing in fourteen dimensions. Then, these branes are compactified on three circles via two different ways (symmetrically
  and anti-symmetrically), and two bosonic and fermionic parts of action for M0-branes are produced. By joining M0-branes, supersymmetric Mp-branes are created which contain the equal number of degrees of freedom for fermions and bosons. Our universe is constructed on one of Mp-branes and other Mp-brane and extra energy play the role of bulk. By dissolving extra energy which is produced by compacting
 actions of Gp-branes, into our universe, the number of degrees of freedom on it and also it's scale factor increase and universe expands. We test G-theory with observations and find that the the magnitude of the slow-roll parameters and the tensor-to-scalar ratio in this model are very smaller than one which are in agreement with predictions of experimental data. Finally, we  consider the origin of the extended theories of gravity in G-theory and show that these theories could be anomaly free.   \\\\


 \end{abstract}
 \date{\today}

\maketitle
\section{Introduction}

Newly, Padmanabhan has investigated the origin of the universe expansion \cite{w1}. He has argued that the expansion
of the universe happens as a result of a difference between the
surface degrees of freedom on the holographic horizon and the bulk
degrees of freedom \cite{w1}. Until now, several authors studied
this interesting idea and its implications for cosmology
\cite{w2,w3,w4,w5,w6,w7}. For example, some authors have applied the Padmanabhan
proposal  to derive the Friedmann equations of an ($n
+ 1$)-dimensional Friedmann-Robertson-Walker (FRW) universe in the
framework of general relativity, Gauss-Bonnet gravity and Lovelock
gravity  \cite{w2}. In another investigation, the Padmanabhan idea has been
generalized to brane cosmology, scalar-tensor cosmology and $f(R)$
gravity \cite{w3}. In another paper, with the help of
this suggestion, authors have calculated the Friedmann
equations of universe  in higher dimensional space-time in
different gravities like Gauss-Bonnet and Lovelock gravity with
general spacial curvature  \cite{w4}. In other consideration,
the Padmanabhan proposal has been generalized to the non-flat
space, relating to the spatial curvature parameter $k = \pm 1$ \cite{w5,w6}. Besides, in
\cite{w7},  the Padmanabhan idea has been studied in the
context of Generalized Uncertainty Principle (GUP). Now, a natural arising question is how  the
Padmanabhan proposal could explain the process of producing  degrees of freedom in the universe and in the bulk.  We
try to answer this question in G-theory which is more complete than M-theory and reduces to it in some limitations.\\
There are various types of string theory like type IIA, type IIB, type I, heterotic (SO(32)) and $E_{8}\times E_{8}$ that each of them could explain some parts of phenomenological events in cosmology, particle physics and other fields of physics \cite{m1,m1t1,m1t2}. These theories discuss about ten dimensional space-time, however the maximum dimension permitted
by supersymmetry of the elementary particles is eleven. In 1987, it has been discussed that  the
Type IIA string may be  the limiting case of the eleven-dimensional supermem-
brane \cite{m2} and in 1994, it was argued that the spectrum of states that
are produced by compactifying the membrane theory from eleven dimensions to four
are similar to those that obtained by compactifying the Type IIA string from ten
dimensions to four \cite{m3,m4}. In 1995, Edward Witten has proposed a new theory, named M-theory whose low energy effective field theory description is 11-dimensional supergravity \cite{t1}. This theory can be reduced to various string theories by compactification and then by various string dualities. However, the algebra that should be applied for M2-branes was unclear. Less than ten years ago, some authors suggested a formalism for M2-branes and showed that for defining the action with $N=8$ supersymmetry (which is an accepted supersymmetry in eleven dimensions), the Lie-3-algebra is needed  \cite{q17,q19,q21,q22,q23}. They introduced two form gauge fields, spinors and scalars in M-theory by using this algebra and obtained the relations between them. \\
Now, the question arises that how M-theory construct Padmanabhan model and explain evolution of degrees of freedom on the surface of universe and in a bulk?   To answer this question,  it has been shown that before universe birth,  there  exist some  M0-branes which are zero dimensional objects and only scalar fields are attached to them without presence of any gauge field and fermions \cite{t7,t8,t9,t10,t11}. Then, the $M0$-branes  glue and 
build a system of $M1$ and anti-$M1$-branes connected by a wormhole which named $M1$-BIon \cite{t7,t8,t9}.  When the M0-branes link  to each other 
symmetrically, gauge 
fields are created and by joining $M0$-branes anti-symmetrically, such as the upper and lower of $M1$-branes became different, fermions are created \cite{t8}. Finally, these $M1$-BIons join to each other and construct
$M3$-BIons, which consists of a configuration of an $M3$, an anti-$M3$-brane connected by a wormhole 
\cite{t7,t8,t9}. Our universe is formed on one of these M3-branes and by growing them, expands   \cite{t7,t8,t9,t10,t11}. In this model, the number of degrees of freedom on the surface of universe depends on the energy of M3-brane and the number of degrees of freedom in a bulk is related to the energy of wormhole and another M3-brane. By dissolving wormhole in our M3-brane, the number of degrees of freedom on the universe surface increases and universe expands.  \\ However, this model is not in good agreement with M-theory. Because, in this theory, only M2 and M5 are stable objects and M3-brane may be produced for a short time. On the other hand, we can't construct our four dimensional universe on M2 and M5 branes which are three and six dimensional objects and their properties are different from  observed properties of universe. Also, some important questions remain without any response in M-theory. For example, what is the origin of creation of different fields like bosonic and fermionic fields in eleven dimensional space-time? \\
To reply to these questions, we have to assume for the existence of extra dimensions in additional to eleven. We have examined different numbers of dimensions and find that if world has fourteen dimensions which three of them be compactified on three circles, supersymmetry emerges without adding any thing by hand. In this theory that we name it G-theory, at the beginning, there isn't any fermion or gauge field and there are only scalars that construct initial G0-branes. If G0-branes are compactified on a circle symmetrically, bosons are produced and by compactifying non-symmetrically, fermions are created. Also, by joining G0-branes and formation of higher dimensional G-branes, different types of gauge fields are born. Finally, by compactifying G-theory on three circles via two ways, one symmetrically and one anti-symmetrically, M-theory and it's related supersymmetry is created. After creation of Mp-branes, our universe is produced on one of them and interact with extra energy 
 and another branes. In these conditions, the energy of this Mp-brane leads to the emergence of degrees of freedom on the universe and the energy of other Mp-brane and extra energy produce degrees of freedom in a bulk. Extra energy dissolves into our universe and leads to an increase in number of degrees of freedom on it and expansion. In fact, this extra energy leads to the inequality between degrees of freedom on the surface and in a bulk and thus, the Padmanabhan mechanism is in good agreement with  G-theory.
\\   
The outline of the paper is as follows. In section \ref{o1}, we will construct G-theory in fourteen dimensions and produce the fermionic fields. In section \ref{o2}, we will show that by compactification of G-theory on three circles, supersymmetric M-theory is emerged. In section \ref{o3}, we will consider the emergence of degrees of freedom on the universe and in a bulk in G-theory and show that the Padmanabhan idea is in good agreement with G-theory. In section \ref{o3}, we will examine the model against observations.  In section \ref{o4}, we will show that extended theories of gravity in G-theory could be anomaly free. The last section is devoted to a summary and conclusion. The last section is devoted to a summary and conclusion. \\

The units used throughout the paper are: $\hbar=c=8\pi G=1$.

\section{The emergence of fermionic and bosonic degrees of freedom and supersymmetry in G-theory }\label{o1}
Before discussing Padmanabhan mechanism in G-theory, we have to consider the  process of the appearance of supersymmetry which has the main role in evolution of degrees of freedom on the universe surface and in a bulk. In G-theory, we assume that at first, there are only G0-branes in fourteen dimensions which only scalars are attached to them without any gauge, bosons and fermionic fields. Second, by  joining G0-branes, G1-branes are emerged and one form of gauge fields are produced. Third, by joining G1-branes, G2-branes are emerged and two form gauge fields are created. Forth, by joining G2-branes, G3-branes are produced and three form gauge bosons are born. Fifth, by joining G3-branes, G4-branes are born and four form gauge fields are created. Sixth, by joining G4-branes, G5-branes are emerged and five form gauge fields are produced. Seventh, these branes join to each other and construct G6-branes which six form gauge fields live on them. In this theory, two form gauge fields play the role of graviton and four form fields play the role of curvature. Also, by compacting G-branes on circles via two different ways (symmetrically and anti-symmetrically), their symmetry is broken and both groups of fermions and bosons  emerge.  

Previously, using Lie-three algebra, the action of M0-brane in M-theory has been given by \cite{t7,t8,t9,t10,t11,q17,q19,q21,q22}:

\begin{eqnarray}
S_{M0} =
T_{M0}\int dt Tr(
\Sigma_{M,N,L=0}^{10}
\langle[X^{M},X^{N},X^{L}],[X^{M},X^{N},X^{L}]\rangle)
\label{at29}
\end{eqnarray}

where $X^{M}=X^{M}_{\alpha}\tilde{T}^{\alpha}$ and

\begin{eqnarray}
 &&[\tilde{T}^{\alpha}, \tilde{T}^{\beta}, \tilde{T}^{\gamma}]= f^{\alpha \beta \gamma}_{\eta}\tilde{T}^{\eta} \nonumber \\&&\langle \tilde{T}^{\alpha}, \tilde{T}^{\beta} \rangle = h^{\alpha\beta} \nonumber \\&& [X^{M},X^{N},X^{L}]=[X^{M}_{\alpha}\tilde{T}^{\alpha},X^{N}_{\beta}\tilde{T}^{\beta},X^{L}_{\gamma}\tilde{T}^{\gamma}]\nonumber \\&&\langle X^{M},X^{M}\rangle = X^{M}_{\alpha}X^{M}_{\beta}\langle \tilde{T}^{\alpha}, \tilde{T}^{\beta} \rangle
\label{att29}
\end{eqnarray}

where  $X^{M}$(i=1,3,...10) are transverse scalars to M0-brane and $\tilde{T}^{\gamma}$ is the generator of Lie-three algebra. By compacting M-theory on eleven dimension, this action transits to the action of D0-brane \cite{q12,q13,q14,q15,qq15,q16}:

\begin{eqnarray}
&& S_{D0} =
-T_{D0}
\int dt Tr(
\Sigma_{m=0}^{9}
[X^{m},X^{n}]^{2})
\label{a15}
\end{eqnarray}

Here $T_{D0}$ is the brane tension and $X^{m}$ are transverse
scalars.

 Now, we introduce the Born-Infeld action for G0-brane by replacing three dimensional
Nambu-Poisson bracket \cite{q17,q19,q21,q22} for Mp-branes by
Six one in action and using the Li-6-algebra \cite{q23}:

\begin{eqnarray}
S_{G0} =
T_{G0}\int dt Tr(
\Sigma_{M,N,L,I,J,K=0}^{13}
\langle[X^{M},X^{N},X^{L},X^{I},X^{J},X^{K}],[X^{M},X^{N},X^{L},X^{I},X^{J},X^{K}]\rangle)
\label{a29}
\end{eqnarray}

where $X^{M}=X^{M}_{\alpha}T^{\alpha}$ and

\begin{eqnarray}
 &&[T^{\alpha}, T^{\beta}, T^{\gamma}, T^{\alpha'}, T^{\beta'}, T^{\gamma'}]= f^{\alpha \beta \gamma \alpha' \beta' \gamma'}_{\eta}T^{\eta} \nonumber \\&&\langle T^{\alpha}, T^{\beta} \rangle = h^{\alpha\beta} \nonumber \\&& [X^{M},X^{N},X^{L}]=[X^{M}_{\alpha}T^{\alpha},X^{N}_{\beta}T^{\beta},X^{L}_{\gamma}T^{\gamma},X^{M'}_{\alpha'}T^{\alpha'},X^{N'}_{\beta'}T^{\beta'},X^{L}_{\gamma'}T^{\gamma'}]\nonumber \\&&\langle X^{M},X^{M}\rangle = X^{M}_{\alpha}X^{M}_{\beta}\langle T^{\alpha}, T^{\beta} \rangle
\label{aa29}
\end{eqnarray}

where  $X^{M}$(i=0,3,...13) are transverse scalars to G0-brane and $T^{\alpha}$ is the generator of group. This action is similar to the action of M0-branes in 11 dimensions and the action of D0-branes in 10 dimensions. By compactifying G-theory on four circles of radius R,
this action will be made a transition into two dimensional action for D0-brane. To show this, we use of the method in \cite{q22}
and define $<X^{10}>=<X^{11}>=<X^{12}>=<X^{13}>=\frac{R}{l_{p}^{3/2}}$ where $l_{p}$ is the Planck length. We have:

\begin{eqnarray}
&& S_{G0} = -
T_{G0}\int dt Tr(
\Sigma_{M,N,L,M',N',L'=0}^{13}
\langle[X^{M},X^{N},X^{L},X^{M'},X^{N'},X^{L'}],[X^{M},X^{N},X^{L},X^{M'},X^{N'},X^{L'}]\rangle)) = \nonumber \\
&& - T_{G0}\int dt Tr(\Sigma_{M,N,L,E,F,G,M',N',L',E',F',G'=0}^{13}\varepsilon_{M,N,L,M',N',L',D}\varepsilon_{E,F,G,E',F',G'}^{D}\times \nonumber \\
&& X^{M}X^{N}X^{L}X^{M'}X^{N'}X^{L'}X^{E}X^{F}X^{G}X^{E'}X^{F'}X^{G'} = \nonumber \\
&& - 6T_{G0}\int dt Tr(\Sigma_{M,N,L,E,F,G,M',N',L',E',F',G'=0}^{12}\varepsilon_{M,N,10,11,12,13,D}\varepsilon_{E,F,10,11,12,13}^{D}\times \nonumber \\
&& X^{M}  X^{N}X^{10}X^{11}X^{12}X^{13}X^{E}X^{F}X^{10}X^{11}X^{12}X^{13} - \nonumber \\
&& 6T_{G0} \int dt \Sigma_{M,N,L,E,F,G,M',N',L',E',F',G'=0,\neq 10,11,12,13}^{9}\varepsilon_{M,N,L,M',N',L',D}\varepsilon_{E,F,G,E',F',G'}^{D}\times\nonumber \\
&& X^{M}X^{N}X^{L}X^{M'}X^{N'}X^{L'}X^{E}X^{F}X^{G}X^{E'}X^{F'}X^{G'} = \nonumber \\
&& - 6T_{G0}(\frac{R^{8}}{l_{p}^{81}})\int dt Tr(\Sigma_{M,N,E,F=0}^{9}\varepsilon_{M,N,10,11,12,13D}\varepsilon_{E,F,10,11,12,13}^{D}X^{M}X^{N}X^{E}X^{F} - \nonumber \\
&&  6T_{G0} \int dt \Sigma_{M,N,L,E,F,G,M',N',L',E',F',G'=0,\neq 10,11,12,13}^{9}\varepsilon_{M,N,L,M',N',L',D}\varepsilon_{E,F,G,E',F',G'}^{D}\times\nonumber \\
&& X^{M}X^{N}X^{L}X^{M'}X^{N'}X^{L'}X^{E}X^{F}X^{G}X^{E'}X^{F'}X^{G'}= \nonumber \\
&& - 6T_{G0}(\frac{R^{8}}{l_{p}^{81}})\int dt Tr(\Sigma_{M,N=0}^{9}[X^{M},X^{N}]^{2}) - \nonumber \\
&&  6T_{G0} \int dt \Sigma_{M,N,L,E,F,G,M',N',L',E',F',G'=0,\neq 10,11,12,13}^{9}\varepsilon_{M,N,L,M',N',L',D}\varepsilon_{E,F,G,E',F',G'}^{D}\times\nonumber \\
&& X^{M}X^{N}X^{L}X^{M'}X^{N'}X^{L'}X^{E}X^{F}X^{G}X^{E'}X^{F'}X^{G'}= \nonumber \\
&& S_{D0} -   6T_{G0} \int dt \Sigma_{M,N,L,E,F,G,M',N',L',E',F',G'=0,\neq 10,11,12,13}^{9}\varepsilon_{M,N,L,M',N',L',D}\varepsilon_{E,F,G,E',F',G'}^{D}\times\nonumber \\
&& X^{M}X^{N}X^{L}X^{M'}X^{N'}X^{L'}X^{E}X^{F}X^{G}X^{E'}X^{F'}X^{G'}=\nonumber \\
&& S_{D0} + V_{Extra,1}
\label{a30}
\end{eqnarray}

where  $T_{G0/D0}$  is tension of brane and $ V_{Extra,1}$ is the extra energy which becomes free during compactification. We define
$T_{D0}=6T_{G0}(\frac{R^{8}}{l_{p}^{81}})=\frac{1}{g_{s}l_{s}}$
where $g_{s}$ and $l_{s}$ are the string coupling and string
length respectively. Thus, the actions in string theory and
G-theory are completely related and all results in string theory
can be generalized to G-theory.

Similar to Dp-branes, different Gp-branes can be built from
G0-brane by using the following rules  \cite{q17,q19,q21,q22}:

\begin{eqnarray}
&&\langle[X^{a},X^{b},X^{a'},X^{b'},X^{c},X^{i}],[X^{a},X^{b},X^{a'},X^{b'},X^{c},X^{i}]\rangle= \nonumber \\&& -
\frac{1}{2}\varepsilon^{abcdef}\varepsilon^{ab'c'd'e'f'}(\partial_{b}\partial_{c}\partial_{d}\partial_{e}\partial_{f}X^{i}_{\alpha})(\partial_{b'}\partial_{c'}\partial_{d'}\partial_{e'}\partial_{f'}X^{i}_{\beta})\langle(T^{\alpha},T^{\beta}\rangle =  \nonumber \\
&&
 -\frac{1}{2}\langle \partial_{b}\partial_{c}\partial_{d}\partial_{e}\partial_{f}X^{i},\partial_{b}\partial_{c}\partial_{d}\partial_{e}\partial_{f}X^{i}\rangle \nonumber \\
 && -\langle[X^{j},X^{b},X^{a'},X^{b'},X^{c},X^{i}],[X^{j},X^{b},X^{a'},X^{b'},X^{c},X^{i}]\rangle= \nonumber \\
 && -\frac{1}{2}\sum_{j}(X^{j})^{2}\varepsilon^{jbcdef}\varepsilon^{jbc'd'e'f'}(\partial_{c}\partial_{d}\partial_{e}\partial_{f'}X^{i}_{\alpha})(\partial_{c'}\partial_{d'}\partial_{e'}\partial_{f'}X^{i}_{\beta})\langle(T^{\alpha},T^{\beta}\rangle =  \nonumber \\
 &&
 -\frac{1}{2}\sum_{j}(X^{j})^{2}\langle \partial_{c}\partial_{d}\partial_{e}\partial_{f}X^{i},\partial_{c}\partial_{d}\partial_{e}\partial_{f}X^{i}\rangle \nonumber \\
   && \langle[X^{j},X^{k},X^{a'},X^{b'},X^{c},X^{i}],[X^{j},X^{k},X^{a'},X^{b'},X^{c},X^{i}]\rangle= \nonumber \\
   && -\frac{1}{2}\sum_{j}(X^{j})^{4}\varepsilon^{jkcdef}\varepsilon^{jkcd'e'f'}(\partial_{d}\partial_{e}\partial_{f}X^{i}_{\alpha})(\partial_{d'}\partial_{e'}\partial_{f'}X^{i}_{\beta})\langle(T^{\alpha},T^{\beta}\rangle =  \nonumber \\
   &&
   - \frac{1}{2}\sum_{j}(X^{j})^{4}\langle \partial_{d}\partial_{e}\partial_{f}X^{i},\partial_{d}\partial_{e}\partial_{f}X^{i}\rangle \nonumber \\
       && \langle[X^{j},X^{k},X^{l},X^{b'},X^{c},X^{i}],[X^{j},X^{k},X^{l},X^{b'},X^{c},X^{i}]\rangle= \nonumber \\
       &&- \frac{1}{2}\sum_{j}(X^{j})^{6}\varepsilon^{jkldef}\varepsilon^{jkld'e'f'}(\partial_{d}\partial_{e}\partial_{f'}X^{i}_{\alpha})(\partial_{d'}\partial_{e'}\partial_{f'}X^{i}_{\beta})\langle(T^{\alpha},T^{\beta}\rangle =  \nonumber \\
       && -
        \frac{1}{2}\sum_{j}(X^{j})^{6}\langle \partial_{e}\partial_{f}X^{i},\partial_{e}\partial_{f}X^{i}\rangle \nonumber \\
               && \langle[X^{j},X^{k},X^{l},X^{m},X^{e},X^{i}],[X^{j},X^{k},X^{l},X^{m},X^{e},X^{i}]\rangle= \nonumber \\
               &&- \frac{1}{2}\sum_{j}(X^{j})^{8}\varepsilon^{jklmne}\varepsilon^{jklmne'}(\partial_{e}X^{i}_{\alpha})(\partial_{e'}X^{i}_{\beta})\langle(T^{\alpha},T^{\beta}\rangle =  \nonumber \\
               &&-
                \frac{1}{2}\sum_{j}(X^{j})^{8}\langle \partial_{e}X^{i},\partial_{e}X^{i}\rangle\nonumber \\
&&\nonumber \\
&&\langle[X^{a},X^{b},X^{c},X^{a'},X^{b'},X^{c'}],[X^{a},X^{b},X^{c},X^{a'},X^{b'},X^{c'}]\rangle=\nonumber \\
&&-
(F^{abca'b'c'}_{\alpha\beta\gamma \alpha'\beta'\gamma'})(F^{abca'b'c'}_{\alpha\beta\eta\alpha'\beta'\eta'})\langle[T^{\alpha},T^{\beta},T^{\gamma},T^{\alpha'},T^{\beta'},T^{\gamma'}],[T^{\alpha},T^{\beta},T^{\eta},T^{\alpha'},T^{\beta'},T^{\eta'}]\rangle)=\nonumber \\
&&- (F^{abca'b'c'}_{\alpha\beta\gamma\alpha'\beta'\gamma'})(F^{abca'b'c'}_{\alpha\beta\eta\alpha'\beta'\eta'})f^{\alpha \beta \gamma \alpha'\beta' \gamma'}_{\sigma}h^{\sigma \kappa}f^{\alpha \beta \eta \alpha' \beta' \eta'}_{\kappa} \langle T^{\gamma},T^{\eta}\rangle= \nonumber \\
&&-
(F^{abca'b'c'}_{\alpha\beta\gamma\alpha'\beta'\gamma'})(F^{abca'b'c'}_{\alpha\beta\eta \alpha'\beta'\eta'})\delta^{\kappa \sigma}\delta^{\alpha\alpha'}\delta^{\beta\beta'}\delta^{\gamma\gamma'}  \langle T^{\gamma},T^{\eta}\rangle=
\langle F^{abca'b'c'},F^{abca'b'c'}\rangle \nonumber \\
&&\langle[X^{i},X^{b},X^{c},X^{a'},X^{b'},X^{c'}],[X^{i},X^{b},X^{c},X^{a'},X^{b'},X^{c'}]\rangle=\nonumber \\
&&-
\sum_{i}(X^{i})^{2} \langle F^{bca'b'c'},F^{bca'b'c'}\rangle  \nonumber \\
&&\langle[X^{i},X^{j},X^{c},X^{a'},X^{b'},X^{c'}],[X^{i},X^{j},X^{c},X^{a'},X^{b'},X^{c'}]\rangle=\nonumber \\
&&-
\sum_{i}(X^{i})^{4} \langle F^{ca'b'c'},F^{ca'b'c'}\rangle \nonumber \\
&&\langle[X^{i},X^{b},X^{c},X^{a'},X^{b'},X^{c'}],[X^{i},X^{b},X^{c},X^{a'},X^{b'},X^{c'}]\rangle=\nonumber \\
&&-
\sum_{i}(X^{i})^{2} \langle F^{bca'b'c'},F^{bca'b'c'}\rangle  \nonumber \\
&&\langle[X^{i},X^{j},X^{k},X^{a'},X^{b'},X^{c'}],[X^{i},X^{j},X^{k},X^{a'},X^{b'},X^{c'}]\rangle=\nonumber \\
&&-
\sum_{i}(X^{i})^{6} \langle F^{a'b'c'},F^{a'b'c'}\rangle  \nonumber \\
&&\langle[X^{i},X^{j},X^{k},X^{l},X^{b'},X^{c'}],[X^{i},X^{j},X^{k},X^{l},X^{b'},X^{c'}]\rangle=\nonumber \\
&&-
\sum_{i}(X^{i})^{8} \langle F^{b'c'},F^{b'c'}\rangle   \nonumber \\
&&\nonumber \\
&&i,j=p+1,..,13\quad a,b=0,1,...p\quad M,N=0,..,13~~
\label{a31}
\end{eqnarray}

where

\begin{eqnarray}
&&F_{abca'b'c'}=\partial_{[a} A_{bca'b'c']}=\partial_{a} A_{bca'b'c'}-\partial_{c'} A_{abca'b'}+\partial_{b'} A_{c'abca'}-...\label{a32}
\end{eqnarray}

\begin{eqnarray}
&&F_{bca'b'c'}=\partial_{[b} A_{ca'b'c']}=\partial_{b} A_{ca'b'c'}-\partial_{c'} A_{bca'b'}+\partial_{b'} A_{c'bca'}-...\label{aq32}
\end{eqnarray}

\begin{eqnarray}
&&F_{ca'b'c'}=\partial_{[c} A_{a'b'c']}=\partial_{c} A_{a'b'c'}-\partial_{c'} A_{ca'b'}+\partial_{b'} A_{c'ca'}-...\label{aqb32}
\end{eqnarray}

\begin{eqnarray}
&&F_{a'b'c'}=\partial_{[a'} A_{b'c']}=\partial_{a'} A_{b'c'}-\partial_{c'} A_{a'b'}+\partial_{b'} A_{c'a'}\label{aw32}
\end{eqnarray}

\begin{eqnarray}
&&F_{b'c'}=\partial_{[b'} A_{c']}=\partial_{b'} A_{c'}-\partial_{c'} A_{b'} \label{aaw32}
\end{eqnarray}

and $A_{bca'b'c'}$ is 5-form guage field, $A_{ca'b'c'}$ is 4-form guage field,  $A_{a'b'c'}$ is 3-form guage field, $A_{a'b'}$ is 2-form guage field and $A_{a}$ is one-form guage field. 

To obtain total action of a p-dimensional system, we should sum over actions of all G0-branes and  use of following action:

\begin{eqnarray}
&& S_{Gp} = \int  d^{p+1}x \sum_{n=1}^{p}\beta_{n}\Big(
\delta^{a_{1},a_{2}...a_{n}}_{b_{1}b_{2}....b_{n}}L^{b_{1}}_{a_{1}}...L^{b_{n}}_{a_{n}}\Big)^{1/2}.\nonumber\\&&
(L)^{a_{n}}_{b_{n}}=\delta_{a_{n}}^{b_{n}} Tr(  \Sigma_{a,b=0}^{p}\Sigma_{j=p+1}^{13}\Big(
\langle[X^{a},X^{b},X^{c},X^{a'},X^{b'},X^{c'}],\langle[X^{a},X^{b},X^{c},X^{a'},X^{b'},X^{c'}]\rangle +\nonumber\\&& 
\langle[X^{a},X^{b},X^{c},X^{a'},X^{b'},X^{i}],\langle[X^{a},X^{b},X^{c},X^{a'},X^{b'},X^{i}]\rangle +\nonumber\\&& 
\langle[X^{j},X^{b},X^{c},X^{a'},X^{b'},X^{i}],\langle[X^{j},X^{b},X^{c},X^{a'},X^{b'},X^{i}]\rangle +\nonumber\\&& 
\langle[X^{j},X^{k},X^{c},X^{a'},X^{b'},X^{i}],\langle[X^{j},X^{k},X^{c},X^{a'},X^{b'},X^{i}]\rangle +\nonumber\\&& 
\langle[X^{j},X^{k},X^{l},X^{a'},X^{b'},X^{i}],\langle[X^{j},X^{k},X^{l},X^{a'},X^{b'},X^{i}]\rangle+\nonumber\\&& 
\langle[X^{j},X^{k},X^{l},X^{m},X^{b'},X^{i}],\langle[X^{j},X^{k},X^{l},X^{m},X^{b'},X^{i}]\rangle +\nonumber\\&& 
\langle[X^{i},X^{b},X^{c},X^{a'},X^{b'},X^{c'}],\langle[X^{i},X^{b},X^{c},X^{a'},X^{b'},X^{c'}]\rangle +\nonumber\\&& 
\langle[X^{i},X^{j},X^{c},X^{a'},X^{b'},X^{c'}],\langle[X^{i},X^{j},X^{c},X^{a'},X^{b'},X^{c'}]\rangle  +\nonumber\\&& 
\langle[X^{i},X^{j},X^{k},X^{a'},X^{b'},X^{c'}],\langle[X^{i},X^{j},X^{k},X^{a'},X^{b'},X^{c'}]\rangle +\nonumber\\&& 
\langle[X^{i},X^{j},X^{k},X^{l},X^{b'},X^{c'}],\langle[X^{i},X^{j},X^{k},X^{l},X^{b'},X^{c'}]\rangle +\nonumber\\&& 
\langle[X^{i},X^{j},X^{k},X^{l},X^{m},X^{n}],\langle[X^{j},X^{k},X^{l},X^{m},X^{n},X^{i}]\rangle
\Big)) \label{s3}
\end{eqnarray}

 Replacing  commutation relations
by derivatives and fields of equations (\ref{a31})
in action (\ref{s3}), we can obtain the relevant action for Gp-brane

\begin{eqnarray}
&& S_{Gp} = \int  d^{p+1}x \sum_{n=1}^{p}\beta_{n}\Big(
\delta^{a_{1},a_{2}...a_{n}}_{b_{1}b_{2}....b_{n}}L^{b_{1}}_{a_{1}}...L^{b_{n}}_{a_{n}}\Big)^{1/2}.\nonumber\\&&
(L)^{a_{n}}_{b_{n}}=\delta_{a_{n}}^{b_{n}} Tr(  \Sigma_{a,b=0}^{p}\Sigma_{j=p+1}^{13}\Big(
\{\frac{1}{2}\langle \partial_{b}\partial_{c}\partial_{d}\partial_{e}\partial_{f}X^{i},\partial_{b}\partial_{c}\partial_{d}\partial_{e}\partial_{f}X^{i}\rangle + \frac{1}{2}\sum_{j}(X^{j})^{2}\langle \partial_{c}\partial_{d}\partial_{e}\partial_{f}X^{i},\partial_{c}\partial_{d}\partial_{e}\partial_{f}X^{i}\rangle + \nonumber \\
&&    \frac{1}{2}\sum_{j}(X^{j})^{4}\langle \partial_{d}\partial_{e}\partial_{f}X^{i},\partial_{d}\partial_{e}\partial_{f}X^{i}\rangle+\frac{1}{2}\sum_{j}(X^{j})^{6}\langle \partial_{e}\partial_{f}X^{i},\partial_{e}\partial_{f}X^{i}\rangle  + \frac{1}{2}\sum_{j}(X^{j})^{8}\langle \partial_{f}X^{i},\partial_{f}X^{i}\rangle + \nonumber \\
&&\frac{1}{720}
\langle F^{abca'b'c'},F^{abca'b'c'}\rangle+\frac{1}{120}\sum_{j}(X^{j})^{2}
\langle F^{abca'b'},F^{abca'b'}\rangle+\frac{1}{24}\sum_{j}(X^{j})^{4}
\langle F^{abca'},F^{abca'}\rangle + \nonumber \\
&&\frac{1}{6}\sum_{j}(X^{j})^{6}
\langle F^{abc},F^{abc}\rangle  +\frac{1}{2}\sum_{j}(X^{j})^{8}
\langle F^{ab},F^{ab}\rangle -\frac{1}{720}
\langle[X^{i},X^{j},X^{k},X^{i'},X^{j'},X^{k'}],[X^{i},X^{j},X^{k},X^{i'},X^{j'},X^{k'}]\rangle
\})
\label{a33}
\end{eqnarray}

 Until now, we have obtained the general action of Gp-branes from G0-branes. This action is not complete, because we have ignored the role of fermionic fields in it. In fact, according to supersymmetric law, we should have  the same number of degrees of freedom for bosons and gauge fields. To produce suppersymmetry, G0-branes is compactified on a circle with two different ways, in one way, strings are compactified completely and bosons are created again and in another manner, strings are compactified non-completely and fermions are emerged. In fact, by breaking the symmetry, scalar strings decay to fermionic strings. To show this, we define  $X \rightarrow \psi^{U}\psi^{L}$ where $\psi^{U/L}$ dentoes the fermionic strings  that are compactified on upper and lower parts of a circle. Also, we use $<X^{G=13}>=\frac{R}{l_{p}^{3/2}}T^{G}$ and $<X^{G'=13}>=\frac{R}{l_{p}^{3/2}}T^{G'}$ for bosons and $<\psi^{L,G=13}>=T^{L,G}\frac{R^{1/2}}{l_{p}^{3/4}}$, $<\psi^{U,G=13}>=T^{U,G}\frac{R^{1/2}
 }{l_{p}^{3/4}}$ and $<X^{G=13}>=\frac{R^{1/2}}{l_{p}^{3/2}}T^{G}$ for fermions and also $\gamma^{M}=T^{L,G}T^{M}$, $T^{M}=T^{L,G}T^{U,G}$ to obtain the same cofficients for both types of fields. We have:

 \begin{eqnarray}
 && S_{G0} = -
 T_{G0}\int dt Tr(
 \Sigma_{M,N,L,M',N',K'=0}^{13}
 \langle[X^{M},X^{N},X^{K},X^{M'},X^{N'},X^{K'}],[X^{M},X^{N},X^{K},X^{M'},X^{N'},X^{L'}]\rangle)) = \nonumber \\
 && - T_{G0}\int dt Tr(\Sigma_{M,N,L,E,F,G,M',N',K',E',F',G'=0}^{13}\varepsilon_{M,N,L,M',N',K',D}\varepsilon_{E,F,G,E',F',G'}^{D}\times \nonumber \\
 && X^{M}X^{N}X^{K}X^{M'}X^{N'}X^{K'}X^{E}X^{F}X^{G}X^{E'}X^{F'}X^{G'} = \nonumber \\
 && - 6T_{G0}\int dt Tr(\Sigma_{M,N,K,E,F,G,M',N',L',E',F',G'=0}^{12}\varepsilon_{M,N,K,E,F,13,D}\varepsilon_{M',N',K',E',F',13}^{D}\times \nonumber \\
 && X^{M}  X^{N}X^{K}X^{E}X^{F}X^{13}X^{E'}X^{F'}X^{M'}X^{N'}X^{K'}X^{13} - \nonumber \\
 && 6T_{G0} \int dt \Sigma_{M,N,L,E,F,G,M',N',K',E',F',G'=0,\neq 13}^{12}\varepsilon_{M,N,K,M',N',L',D}\varepsilon_{E,F,G,E',F',G'}^{D}\times\nonumber \\
 && X^{M}X^{N}X^{K}X^{M'}X^{N'}X^{K'}X^{E}X^{F}X^{G}X^{E'}X^{F'}X^{G'} =  \nonumber \\
  && - 3T_{G0}\int dt Tr(\Sigma_{M,N,K,E,F,G,M',N',K',E',F',G'=0}^{12}\varepsilon_{M,N,K,E,F,13,D}\varepsilon_{M',N',K',E',F',13}^{D}\times \nonumber \\
  && X^{M}  X^{N}X^{K}X^{E}X^{F}\psi^{U,13}\psi^{L,13}X^{E'}X^{F'}X^{M'}X^{N'}X^{K'}\psi^{U,13}\psi^{L,13} \nonumber \\
   && - 3T_{G0}\int dt Tr(\Sigma_{M,N,K,E,F,G,M',N',K',E',F',G'=0}^{12}\varepsilon_{M,N,K,E,F,13,D}\varepsilon_{M',N',K',E',F',13}{D}\times \nonumber \\
   && X^{M}  X^{N}X^{K}X^{E}X^{F}X^{13}X^{E'}X^{F'}X^{M'}X^{N'}X^{K'}X^{13} - \nonumber \\
  && 6T_{G0} \int dt \Sigma_{M,N,K,E,F,G,M',N',K',E',F',G'=0,\neq 12}^{12}\varepsilon_{M,N,K,E,F,G,D}\varepsilon_{M',N',K',E',F',G'}^{D}\times\nonumber \\
   && X^{M}X^{N}X^{K}X^{M'}X^{N'}X^{K'}X^{E}X^{F}X^{G}X^{E'}X^{F'}X^{G'} = \nonumber \\
     && - 3T_{G0}(\frac{R_{p}^{2}}{l_{p}^{3}})\int dt Tr(\Sigma_{M,N,K,E,F,G,M',N',K',E',F',G'=0}^{12}\varepsilon_{M,N,K,E,F,13,D}\varepsilon_{M',N',K',E',F',13}^{D}\times \nonumber \\
     &&  \gamma^{M} X^{N}X^{K}X^{E}X^{F}\psi^{U,13}X^{E'}X^{F'}X^{M'}X^{N'}X^{K'}\psi^{U,13}- \nonumber \\
      &&  3T_{G0}(\frac{R^{2}}{l_{p}^{3}})\int dt Tr(\Sigma_{M,N,L,E,F,G,M',N',K',E',F',G'=0}^{12}\varepsilon_{M,N,K,E,F,13,D}\varepsilon_{M',N',K',E',F',13}^{D}\times \nonumber \\
      && X^{M}  X^{N}X^{K}X^{E}X^{F}X^{E'}X^{F'}X^{M'}X^{N'}X^{K'} - \nonumber \\
     && 6T_{G0} \int dt \Sigma_{M,N,K,E,F,G,M',N',K',E',F',G'=0,\neq 12}^{11}\varepsilon_{M,N,K,E,F,G,D}\varepsilon_{M',N',K',E',F',G'}^{D}\times\nonumber \\
      && X^{M}X^{N}X^{K}X^{M'}X^{N'}X^{K'}X^{E}X^{F}X^{G}X^{E'}X^{F'}X^{G'} =\nonumber \\
       && - 3T_{G0}(\frac{R^{2}}{l_{p}^{3}})\int dt Tr(\Sigma_{M,N,K,E=0}^{12}
        \langle[\gamma^{M},X^{N},X^{K},X^{E},X^{F},\psi^{U,13}],[X^{M},X^{N},X^{K},X^{E},X^{F},\psi^{U,13}]\rangle) \nonumber \\
 && - 3T_{G0}(\frac{R^{2}}{l_{p}^{3}})\int dt Tr(\Sigma_{M,N,K,E=0}^{12}
  \langle[X^{M},X^{N},X^{K},X^{E},X^{F}],[X^{M},X^{N},X^{K},X^{E},X^{F}]\rangle)  - \nonumber \\
       && 6T_{G0} \int dt \Sigma_{M,N,K,E,F,G,M',N',K',E',F',G'=0,\neq 12}^{11}\varepsilon_{M,N,K,E,F,G,D}\varepsilon_{M',N',K',E',F',G'}^{D}\times\nonumber \\
             && X^{M}X^{N}X^{K}X^{M'}X^{N'}X^{K'}X^{E}X^{F}X^{G}X^{E'}X^{F'}X^{G'} = \nonumber \\
 && S_{G0}^{Comp-Fermi}+S_{G0}^{Comp-Bosonic} -   6T_{G0} \int dt \Sigma_{M,N,K,E,F,G,M',N',K',E',F',G'=0,\neq 12}^{11}\varepsilon_{M,N,K,E,F,G,D}\varepsilon_{M',N',K',E',F',G'}^{D}\times \nonumber \\
       && X^{M}X^{N}X^{K}X^{M'}X^{N'}X^{K'}X^{E}X^{F}X^{G}X^{E'}X^{F'}X^{G'}=\nonumber \\
 && S_{G0}^{Comp-Fermi}+S_{G0}^{Comp-Bosonic} + V_{Extra,1}
 \label{aO30}
 \end{eqnarray}

 It is clear that by compacting strings in two different ways (symmetrically
   and anti-symmetrically), two different actions are emerged and two different fields are produced. Also, the cofficients of two actions are the same and N=13 supersymmetry  is created.  In this type of supersymmetry, we have 13 bosons and 13 fermions which are originated from two different types of compacting 14 initial bosons.

 \section{The reduction of G-theory to M-theory}\label{o2}
 
 In this section, we show that by compacting G-theory on three circles and via two different ways (symmetrically
    and anti-symmetrically), supersymmetry is produced which contains the same number of degrees of freedom for both fermions and bosons. To obtain supersymmetry in eleven dimensions, we should use of the mechanism in previous section  three times and compactify G-branes on three other cirles. To do this, we use $<X^{G=13}>=\frac{R}{l_{p}^{3/2}}T^{G=13}$, $<X^{F=12}>=\frac{R}{l_{p}^{3/2}}T^{F=12}$, $<X^{E=11}>=\frac{R}{l_{p}^{3/2}}T^{E=11}$ for bosons and $<\psi^{L,G=13}>=\frac{R^{1/2}}{l_{p}^{3/4}}T^{L,G=13}$, $<\psi^{L,F=12}>=\frac{R^{1/2}}{l_{p}^{3/4}}T^{L,F=12}$, $<\psi^{L,E=11}>=\frac{R^{1/2}}{l_{p}^{3/4}}T^{L,E=11}$ and $<X^{G=13}>=\frac{R}{l_{p}^{3/2}}T^{G=13}$, $<X^{F=12}>=\frac{R}{l_{p}^{3/2}}T^{F=12}$, $<X^{E=11}>=\frac{R}{l_{p}^{3/2}}T^{E=11}$ for fermions in action (\ref{aO30}) to obtain the same cofficients for both types of fields:

  \begin{eqnarray}
  && S_{M0} =
        - 3T_{G0}(\frac{R^{12}}{l_{p}^{27}})\int dt Tr(\Sigma_{M,N,L,E=0}^{10}
         \langle[\gamma^{M},\gamma^{N},\gamma^{L},\psi^{U,11},\psi^{U,12},\psi^{U,13}],[X^{M},X^{N},X^{L},\psi^{U,11},\psi^{U,12},\psi^{U,13}]\rangle) \nonumber \\
  && - 3T_{G0}(\frac{R^{12}}{l_{p}^{27}})\int dt Tr(\Sigma_{M,N,L,E=0}^{10}
   \langle[X^{M},X^{N},X^{L}],[X^{M},X^{N},X^{L}]\rangle)   - \nonumber \\
             && 6T_{G0} \int dt \Sigma_{M,N,K,E,F,G,M',N',K',E',F',G'=0,\neq 11,12,13}^{10}\varepsilon_{M,N,K,E,F,G,D}\varepsilon_{M',N',K',E',F',G'}^{D}\times\nonumber \\
                   && X^{M}X^{N}X^{K}X^{M'}X^{N'}X^{K'}X^{E}X^{F}X^{G}X^{E'}X^{F'}X^{G'} = \nonumber \\
  && S_{M0}^{Fermi}+S_{M0}^{Bosonic}  - \nonumber \\
            && 6T_{G0} \int dt \Sigma_{M,N,K,E,F,G,M',N',K',E',F',G'=0,\neq 11,12,13}^{10}\varepsilon_{M,N,K,E,F,G,D}\varepsilon_{M',N',K',E',F',G'}^{D}\times\nonumber \\
                  && X^{M}X^{N}X^{K}X^{M'}X^{N'}X^{K'}X^{E}X^{F}X^{G}X^{E'}X^{F'}X^{G'} =\nonumber \\
  && S_{M0}^{Fermi}+S_{M0}^{Bosonic} + V_{Extra,1}
  \label{aOP31}
  \end{eqnarray}
 
where we have used of this fact that $T_{M0}=3T_{G0}(\frac{R^{12}}{l_{p}^{27}})$. This equation shows that by compacting G-model on three circles, the N=8 supersymmetric M-theory is emerged which includes the equal number of bosons and fermions. Now, we can show that by joining branes, superpartners are created and Dirac equation can be obtained.  To this end, we use of following laws \cite{q17,q19,q21,q22}:

\begin{eqnarray}
&& \langle[X^{i},X^{b},X^{i}],[X^{i},X^{b},X^{i}]\rangle=
\frac{1}{2}\varepsilon^{abc}\varepsilon^{abd}\Sigma_{i}(X^{i})^{2}(\partial_{a}X^{i}_{\alpha})(\partial_{a}X^{i}_{\beta})\langle(T^{\alpha},T^{\beta}\rangle =
 \frac{1}{2}\Sigma_{i}(X^{i})^{2}\langle \partial_{a}X^{i},\partial_{a}X^{i}\rangle  \nonumber \\
 &&\nonumber \\
&&\langle[X^{a},X^{b},X^{c}],[X^{a},X^{b},X^{c}]\rangle=
(F^{abc}_{\alpha\beta\gamma})(F^{abc}_{\alpha\beta\eta})\langle[T^{\alpha},T^{\beta},T^{\gamma}],[T^{\alpha},T^{\beta},T^{\eta}]\rangle)=\nonumber \\
&& (F^{abc}_{\alpha\beta\gamma})(F^{abc}_{\alpha\beta\eta})f^{\alpha \beta \gamma}_{\sigma}h^{\sigma \kappa}f^{\alpha \beta \eta}_{\kappa} \langle T^{\gamma},T^{\eta}\rangle=
(F^{abc}_{\alpha\beta\gamma})(F^{abc}_{\alpha\beta\eta})\delta^{\kappa \sigma} \langle T^{\gamma},T^{\eta}\rangle=
\langle F^{abc},F^{abc}\rangle \nonumber \\
&& \nonumber \\
&& \langle[\gamma^{a},\gamma^{b},\gamma^{c},\psi^{U,11},\psi^{U,12},\psi^{U,13}],[X^{a},X^{b},X^{c},\psi^{U,11},\psi^{U,12},\psi^{U,13}]\rangle =  \nonumber \\
&& \frac{1}{6}( (\psi^{ \dag U,11}\psi^{ \dag U,12}\psi^{ \dag U,13}) \gamma^{a}\gamma^{b}\gamma^{c}\partial_{[a}\partial_{b}\partial_{c]}(\psi^{ U,11}\psi^{ U,12}\psi^{ U,13})\nonumber \\
&& \nonumber \\
&& \langle[\gamma^{a},\gamma^{b},\gamma^{c},\psi^{U,11},\psi^{U,12},\psi^{U,13}],[X^{i},X^{i},X^{c'},\psi^{U,11},\psi^{U,12},\psi^{U,13}]\rangle =  \nonumber \\
&& \frac{1}{2}\Sigma_{i}(X^{i})^{2} (\psi^{ \dag U,11}\psi^{ \dag U,12}\psi^{ \dag U,13}) \gamma^{a}\gamma^{b}\gamma^{c}\partial_{c'}(\psi^{ U,11}\psi^{ U,12}\psi^{ U,13})=\nonumber \\
&&  \frac{1}{2}\Sigma_{i}(X^{i})^{2} (\psi^{ \dag U,11}\psi^{ \dag U,12}\psi^{ \dag U,13})i\tilde{\gamma}^{c'}\partial_{c'}(\psi^{ U,11}\psi^{ U,12}\psi^{ U,13})\nonumber \\
&&\nonumber \\
&&i\tilde{\gamma}^{c'}= i\tilde{\gamma}^{abc}=\gamma^{a}\gamma^{b}\gamma^{c} \nonumber \\
&&\nonumber \\
&&\Sigma_{m}\rightarrow \frac{1}{(2\pi)^{p}}\int d^{p+1}\sigma \Sigma_{m-p-1}
i,j=p+1,..,10\quad a,b=0,1,...p\quad m,n=0,..,10~~
\label{aa31}
\end{eqnarray}

where

\begin{eqnarray}
&&F_{abc}=\partial_{a} A_{bc}-\partial_{b} A_{ca}+\partial_{c} A_{ab}\label{a32}
\end{eqnarray}

and $A_{ab}$ is 2-form gauge field.  Replacing  commutation relations
by derivatives and fields of equations (\ref{aa31})
in action (\ref{aOP31}), we can obtain the relevant action for Mp-brane

\begin{eqnarray}
&&  S_{Mp} = \Sigma_{a=0}^{p}(S_{M0}^{Fermi}+S_{M0}^{Bosonic})= \nonumber \\ &&-\Sigma_{a=0}^{p}T_{M0}
 \int dt Tr(
 \Sigma_{m=0}^{10}
 \langle[X^{a},X^{b},X^{c}],[X^{a},X^{b},X^{c}]\rangle+  \nonumber \\ && \langle[\gamma^{M},\gamma^{N},\gamma^{L},\psi^{U,11},\psi^{U,12},\psi^{U,13}],[X^{M},X^{N},X^{L},\psi^{U,11},\psi^{U,12},\psi^{U,13}]\rangle) =  \nonumber \\ &&
-T_{Mp} \int d^{p+1}\sigma Tr
(\Sigma_{a,b,c=0}^{p}
\Sigma_{i,j,k=p+1}^{10}
\{\frac{1}{2}\Sigma_{i}(X^{i})^{2} \langle\partial_{a}X^{i},\partial_{a}X^{i}\rangle +\frac{1}{6}
\langle F_{abc},F_{abc}\rangle+
\nonumber \\ &&  \frac{1}{2}\Sigma_{i}(X^{i})^{2} (\psi^{ \dag U,11}\psi^{ \dag U,12}\psi^{ \dag U,13})i\tilde{\gamma}^{c'}\partial_{c'}(\psi^{ U,11}\psi^{ U,12}\psi^{ U,13}) + \nonumber \\ &&  \frac{1}{6}( (\psi^{ \dag U,11}\psi^{ \dag U,12}\psi^{ \dag U,13}) \gamma^{a}\gamma^{b}\gamma^{c}\partial_{[a}\partial_{b}\partial_{c]}(\psi^{ U,11}\psi^{ U,12}\psi^{ U,13}) 
\})
\label{aH33}
\end{eqnarray}

Until now, we have choosen three spacial dimensions (for exmaple, i= 11,12 and 13) and did compactification in that directions. However, we can remove this limitation and assume that compactifications may occur in different directions (i=p+1,...13) which are perpendicular to branes. By replacing $\psi^{ U,11}\psi^{ U,12}\psi^{ U,13}\rightarrow \Psi^{i}$ and choosing $\Sigma_{i}(X^{i})^{2}\rightarrow 1$, we can obtain following action:

 \begin{eqnarray}
  && S_{Mp} = \Sigma_{a=0}^{p}(S_{M0}^{Fermi}+S_{M0}^{Bosonic})= \nonumber \\ &&-\Sigma_{a=0}^{p}T_{M0}
 \int dt Tr(
 \Sigma_{m=0}^{10}
 \langle[X^{a},X^{b},X^{c}],[X^{a},X^{b},X^{c}]\rangle+  \nonumber \\ && \langle[\gamma^{M},\gamma^{N},\gamma^{L},\psi^{U,11},\psi^{U,12},\psi^{U,13}],[X^{M},X^{N},X^{L},\psi^{U,11},\psi^{U,12},\psi^{U,13}]\rangle) =  \nonumber \\ &&
 -T_{Mp} \int d^{p+1}\sigma Tr
 (\Sigma_{a,b,c=0}^{p}
 \Sigma_{i,j,k=p+1}^{10}
 \{\frac{1}{2}\langle\partial_{a}X^{i},\partial_{a}X^{i}\rangle +\frac{1}{6}
 \langle F_{abc},F_{abc}\rangle+
 \nonumber \\ &&  \frac{1}{2} (\Psi^{ \dag U,i})i\tilde{\gamma}^{c'}\partial_{c'}(\Psi^{ U,i}) +  \frac{1}{6}\Psi^{ \dag U,i} i \tilde{\gamma}^{abc}\kappa^{i}\partial_{[a}\chi^{ U}_{bc]} 
 \})
 \label{ap33}
 \end{eqnarray}
 
 where we have defined:

  \begin{eqnarray}
  && \partial_{[a}\chi^{ U}_{bc]}=\partial_{a} \chi^{ U}_{bc}-\partial_{b} \chi^{ U}_{ca}+\partial_{c} \chi^{ U}_{ab} \nonumber \\ && \chi^{ U}_{ab}= \kappa_{i}\partial_{[a}\partial_{b]}(\psi^{ U,i}\psi^{ U,i}\psi^{ U,i}) \nonumber \\ && \kappa^{i}\kappa_{i}=1
  \label{ass33}
  \end{eqnarray}
  
 This action is a generalization of following action with substituting $\beta_{2}=1$ and $\beta_{n}=0$.

  \begin{eqnarray}
   && S_{Mp} =  \int  d^{p+1}x \sum_{n=1}^{p}\beta_{n}\Big(
   \delta^{a_{1},a_{2}...a_{n}}_{b_{1}b_{2}....b_{n}}L^{b_{1}}_{a_{1}}...L^{b_{n}}_{a_{n}}\Big)^{1/2}.\nonumber\\&&
   (L)^{a_{n}}_{b_{n}}=\delta_{a_{n}}^{b_{n}}  Tr
  (\Sigma_{a,b,c=0}^{p}
  \Sigma_{i,j,k=p+1}^{10}
  \{\frac{1}{2}\langle\partial_{a}X^{i},\partial_{a}X^{i}\rangle +\frac{1}{6}
  \langle F_{abc},F_{abc}\rangle+
  \nonumber \\ &&  \frac{1}{2} (\Psi^{ \dag U,i})i\tilde{\gamma}^{c'}\partial_{c'}(\Psi^{ U,i}) +  \frac{1}{6}\Psi^{ \dag U,i} i \tilde{\gamma}^{abc}\kappa^{i}\partial_{[a}\chi^{ U}_{bc]} 
  \})
  \label{apty33}
  \end{eqnarray}
 
This action is in very good agreement with usual action for Mp-branes\cite{q17,q19,q21,q22,q23}. In additional,  first and third terms and also second and four terms are the same which means that there are very good connections between fermions and bosons in this system. In fact, each scalar string $X^{i}$ has a superpartner $\Psi^{ \dag U,i}$ and each two form bosonic vector $A_{ab}$ has a fermionic superpartner $\chi^{ U}_{ab}$. Thus, numbers of degrees of freedom of bosons are equal to fermions and supersymmetry emerges. For example, for M2-brane, indices of scalars and fermions change from 1 to 8 and we will have N=8 supersymmetry as it is predicted by recent papers in \cite{q17,q19,q21,q22,q23}.  The evolutions of both groups of fermions and bosons have the main role in the emergence of inequality between  the number of degrees of freedom on the universe surface and in the bulk and according to the Padmanabhan idea, this inequality leads to expansion of universe. 

\section{Emergence of degrees of freedom and application of Padmanabhan mechanism in G-theory}\label{o3}
 In this section, we show that all dimensions and degrees of freedom can be produced directly from the potential of  scalar strings in G-theory. We assume that at the beginning, two scalar and anti-scalar strings are produced from nothing such as the energy of the scalar string has positive sign like the sign of repulsive potential and the energy of the anti-scalar string has the negative sign like the sign of attractive potential. We have:

  \begin{eqnarray}
  && E \equiv 0\equiv E_{1}+E_{2}\equiv N_{sur}+N_{bulk}\equiv k (X^{14})^{2}- k (X^{14})^{2}
  \label{tm1}
  \end{eqnarray}
  
  where $X^{14}=X^{14}_{\beta}T^{\beta}$. These strings are placed on 14th dimension, however they are excited and produce some degrees of freedom in other 13 dimensions of G-theory. To show this, we rewrite above equations as the follows:

    \begin{eqnarray}
    && E \equiv 0\equiv k \int d^{2p}x \varepsilon^{i_{1}i_{2}...i_{p}14}\varepsilon^{i_{1}i_{2}...i_{p}14} (\frac{\partial}{\partial x_{i_{1}}}\frac{\partial}{\partial x_{i_{2}}}..\frac{\partial}{\partial x_{i_{p}}})^{2}(X^{14})^{2}-\nonumber \\ &&  k \int d^{2p}x \varepsilon^{i_{1}i_{2}...i_{p}14}\varepsilon^{i_{1}i_{2}...i_{p}14} (\frac{\partial}{\partial x_{i_{1}}}\frac{\partial}{\partial x_{i_{2}}}..\frac{\partial}{\partial x_{i_{p}}})^{2}(X^{14})^{2}
    \label{tm2}
    \end{eqnarray}
  
 where we have used of this fact that $\varepsilon^{i_{1}i_{2}...i_{p}}\varepsilon^{i_{1}i_{2}...i_{p}}=1$ and p=13 for 14 dimensional brane in G-model. These new energies are related to new objects in G-theory which are produced by exciting strings. We can show that these are actions of Gp-branes by using following relations between derivatives and brackets \cite{q17,q19,q21,q22,q23,q12,q13,qt13,q14,q15}:

    \begin{eqnarray}
    && \frac{\partial}{\partial x_{i_{1}}}X^{14}=[ X^{i_{1}},X^{14}] \nonumber \\ && \frac{\partial}{\partial x_{i_{1}}}\frac{\partial}{\partial x_{i_{2}}}X^{14}=[ X^{i_{1}},X^{i_{2}},X^{14}]\nonumber \\ && (\frac{\partial}{\partial x_{i_{1}}}\frac{\partial}{\partial x_{i_{2}}}..\frac{\partial}{\partial x_{i_{p}}})(X^{14})=[ X^{i_{1}},X^{i_{2}},...,X^{i_{p}},X^{14}]\nonumber \\ &&\varepsilon^{i_{1}i_{2}...i_{p}14}\varepsilon^{i_{1}i_{2}...i_{p}14} (\frac{\partial}{\partial x_{i_{1}}}\frac{\partial}{\partial x_{i_{2}}}..\frac{\partial}{\partial x_{i_{p}}})^{2}(X^{14})^{2}=\nonumber \\ && \varepsilon^{i_{1}i_{2}...i_{p}14}\varepsilon^{i'_{1}i'_{2}...i'_{p}14} [(\frac{\partial}{\partial x_{i_{1}}}\frac{\partial}{\partial x_{i_{2}}}..\frac{\partial}{\partial x_{i_{p}}})(X^{14})][(\frac{\partial}{\partial x_{i'_{1}}}\frac{\partial}{\partial x_{i'_{2}}}..\frac{\partial}{\partial x_{i'_{p}}})(X^{14})]=\nonumber \\ && \langle [ X_{i_{1}},X_{i_{2}},...,X_{i_{p}},X_{14}],[ X_{i_{1}},X_{i_{2}},
 ...,X_{i_{p}},X_{14}] \rangle
    \label{tm3}
    \end{eqnarray}
  
where $X_{i_{n}}=X_{i_{n}}^{\beta}T_{\beta}$. Using the relations in equation (\ref{tm3}) in equation (\ref{tm2}), we obtain:

    \begin{eqnarray}
    && E \equiv 0\equiv E_{1}+E_{2}\equiv \nonumber \\ && k \int d^{2p}x \langle [ X_{i_{1}},X_{i_{2}},...,X_{i_{p}},X_{14}],[ X_{i_{1}},X_{i_{2}},...,X_{i_{p}},X_{14}] \rangle-\nonumber \\ &&  k \int d^{2p}x \langle [ X_{i_{1}},X_{i_{2}},...,X_{i_{p}},X_{14}],[ X_{i_{1}},X_{i_{2}},...,X_{i_{p}},X_{14}] \rangle
    \label{tm4}
    \end{eqnarray}
    
    This equation shows that excitations of each scalar string in dimension 14 produce various degrees of freedom and new motions of the string and the emergence of other 13 dimensions. These new actions are similar to actions of zero dimensional branes in fourteen dimensions. We can name these new objects as G0-branes. Until now, all dimensions are the same and  there isn't any difference between time and space. Now, by compacting dimensions, we break the symmetry and produce the real space time. To achieve this, we apply the mechanism  in \cite{q22} and define $X_{i_{n=1,3,5,7,9,11,13,14}}=\frac{R}{l_{p}^{3/2}}$   where $l_{p}$ is the Planck length. We get:

    \begin{eqnarray}
    && E_{1}\equiv k \int d^{2p}x \langle [ X_{i_{1}},X_{i_{2}},...,X_{i_{p}},X_{14}],[ X_{i_{1}},X_{i_{2}},...,X_{i_{p}},X_{14}] \rangle=\nonumber \\ &&k \int d^{2p}x   \varepsilon^{i_{1}i_{2}...i_{p}14}\varepsilon^{i'_{1}i'_{2}...i'_{p}14}X_{i_{1}}X_{i_{2}}...X_{i_{p}}X_{14}X_{i'_{1}}X_{i'_{2}}...X_{i'_{p}}X_{14} =\nonumber \\ && k \int d^{2p}x  \varepsilon^{1i_{2}...1314}\varepsilon^{1i'_{2}...1314}X_{i_{2}}...X_{i_{P-1}}X_{14}X_{i'_{2}}...X_{i_{P-1}}X_{14}=\nonumber \\ && -k(\frac{R^{p+1}}{l_{p+1}^{3(p+1)/2}}) \int d^{p}x  \varepsilon^{j_{1}...j_{(p-1)/2}}\varepsilon^{j'_{1}...j'_{(p-1)/2}}X_{j_{1}}...X_{j_{(p-1)/2}}X_{j'_{1}}...X_{j'_{(p-1)/2}}=\nonumber \\ && -k(\frac{R^{p+1}}{l_{p+1}^{3(p+1)/2}}) \int d^{p}x \langle [ X_{j_{1}},X_{j_{2}},...,X_{j_{(p-1)/2}}],[ X_{j_{1}},X_{j_{2}},...,X_{j_{(p-1)/2}}]\rangle=\nonumber \\ && k(\frac{R^{p+1}}{l_{p+1}^{3(p+1)/2}}) \int d^{p}x \langle [i X_{j_{1}},X_{j_{2}},...,X_{j_{(p-1)/2}}],[ iX_{j_{1}},X_{j_{2}},...,X_{j_{(p-1)/2}}]\rangle=\nonumber \\ && k(\frac{R^{p+1}}{l_{p+1}^{3(p+1)/2}}) \int d^{p}x \langle [i t,X_{j_{2}},...,X_{j_{(p-1)/2}}],[ it,X_{j_{2}},...,X_{j_{(p-1)/2}}]\rangle
    \label{tm5}
    \end{eqnarray}  
  where we have used $ \varepsilon^{1i_{2}...1314}\varepsilon^{1i'_{2}...1314}=-\varepsilon^{j_{1}...j_{(p-1)/2}}\varepsilon^{j'_{1}...j'_{(p-1)/2}}$, $X_{j_{1}}=t$ and $i^{2}=-1$.  We can regard the probability for changing the place of strings ($X_{a}$) in this action and rewrite it as follows:

      \begin{eqnarray}
      && E_{1}\equiv T_{G0} \int d^{p}x \Sigma_{a,b,c,d,e,f=0}^{13}\langle [ X_{a},X_{b},X_{c},X_{d},X_{e},X_{f}],[X_{a},X_{b},X_{c},X_{d},X_{e},X_{f}]\rangle
           \label{tmm5}
      \end{eqnarray}  
      
  where we have used $T_{G0}= k(\frac{R^{p+1}}{l_{p+1}^{3(p+1)/2}})$ and $X^{0}=it_{1} $. This equation shows that by compacting some dimensions, the symmetry is broken and one of dimensions can act different from other dimensions which we name it as time. Also, the Lie-6-algebra can be obtained exactly by breaking the symmetry and coincidence with the birth of time for p=13. For another brane, we obtain:

      \begin{eqnarray}
      && E_{2}\equiv - k \int d^{2p}x \langle [ X_{i_{1}},X_{i_{2}},...,X_{i_{p}},X_{14}],[ X_{i_{1}},X_{i_{2}},...,X_{i_{p}},X_{14}] \rangle=\nonumber \\ && -k \int d^{2p}x   \varepsilon^{i_{1}i_{2}...i_{p}14}\varepsilon^{i'_{1}i'_{2}...i'_{p}14}X_{i_{1}}X_{i_{2}}...X_{i_{p}}X_{14}X_{i'_{1}}X_{i'_{2}}...X_{i'_{p}}X_{14} =\nonumber \\ && -k \int d^{2p}x  \varepsilon^{1i_{2}...1314}\varepsilon^{1i'_{2}...1314}X_{i_{2}}...X_{i_{P-1}}X_{14}X_{i'_{2}}...X_{i_{P-1}}X_{14}=\nonumber \\ && k(\frac{R^{p+1}}{l_{p+1}^{3(p+1)/2}}) \int d^{p}x  \varepsilon^{j_{1}...j_{(p-1)/2}}\varepsilon^{j'_{1}...j'_{(p-1)/2}}X_{j_{1}}...X_{j_{(p-1)/2}}X_{j'_{1}}...X_{j'_{(p-1)/2}}=\nonumber \\ && k(\frac{R^{p+1}}{l_{p+1}^{3(p+1)/2}}) \int d^{p}x \langle [ X_{j_{1}},X_{j_{2}},...,X_{j_{(p-1)/2}}],[ X_{j_{1}},X_{j_{2}},...,X_{j_{(p-1)/2}}]\rangle=\nonumber \\ && k(\frac{R^{p+1}}{l_{p+1}^{3(p+1)/2}}) \int d^{p}x \langle [i X_{j_{1}},iX_{j_{2}},...,X_{j_{(p-1)/2}}],[ iX_{j_{1}},iX_{j_{2}},...,X_{j_{(p-1)/2}}]\rangle=\nonumber \\ && k(\frac{R^{p+1}}{l_{p+1}^{3(p+1)/2}}) \int d^{p}x \langle [i t_{1},it_{2},...,X_{j_{(p-1)/2}}],[ it_{1},it_{2},...,X_{j_{(p-1)/2}}]\rangle
      \label{tm6}
      \end{eqnarray} 
      
      where we have assumed $iX_{j_{1}}=it_{1},iX_{j_{2}}=it_{2}$. After regarding the probability for changing the place of strings in this action, we can rewrite it as follows:

            \begin{eqnarray}
            && E_{2}\equiv T_{G0} \int d^{p}x \Sigma_{\tilde{a},\tilde{b},\tilde{c},\tilde{d},\tilde{e},\tilde{f}=0}^{13}\langle [ X_{\tilde{a}},X_{\tilde{b}},X_{\tilde{c}},X_{\tilde{d}},X_{\tilde{e}},X_{\tilde{f}}],[ X_{\tilde{a}},X_{\tilde{b}},X_{\tilde{c}},X_{\tilde{d}},X_{\tilde{e}},X_{\tilde{f}}]\rangle
                 \label{tmm6}
            \end{eqnarray}

      where we have used $T_{G0}= k(\frac{R^{p+1}}{l_{p+1}^{3(p+1)/2}})$ and $X^{0}=it_{1}, X^{1}=it_{2} $. It is clear from above energy that by breaking the symmetry for second brane, two different times are created. In fact, the excitation of an initial string with negative energy causes to production at least two times in action of second G0-brane which makes it's properties different from first G0-brane. This may lead to production of two universes with two physical properties.

            Now, we can use of the method in previous sections and compactify G0-branes on three circles and obtain the energy of Mp-branes, anti-Mp-branes and extra energy. Also, number of degrees of freedom on the universe depends on the energy of Mp-brane and number of degrees of freedom in the bulk is related to the energy of anti-Mp-brane and extra energy. Using (\ref{a31}), (\ref{a33}), (\ref{aOP31}) and (\ref{apty33}), we obtain:

  \begin{eqnarray}
   && N_{sur}\approx E_{Mp} = T_{Mp} \int  d^{p+1}x \Big( Tr
     (\Sigma_{a,b,c=0}^{p}
     \Sigma_{i,j,k=p+1}^{10}
     \{\frac{1}{2}\langle\partial_{a}X^{i},\partial_{a}X^{i}\rangle +\frac{1}{6}
     \langle F_{abc},F_{abc}\rangle+
     \nonumber \\ &&  \frac{1}{2} (\Psi^{ \dag U,i})i\tilde{\gamma}^{c'}\partial_{c'}(\Psi^{ U,i}) +  \frac{1}{6}\Psi^{ \dag U,i} i \tilde{\gamma}^{abc}\kappa^{i}\partial_{[a}\chi^{ U}_{bc]} 
     \})\Big)^{1/2}.
  \label{sc1}
  \end{eqnarray}

    \begin{eqnarray}
     && N_{bulk}\approx E_{anti-Mp}+ E_{extra} \nonumber \\ && \nonumber \\ && E_{anti-Mp} = T_{Mp} \int  d^{p+1}x \Big( Tr
       (\Sigma_{\tilde{a},\tilde{b},\tilde{c}=0}^{p}
       \Sigma_{i,j,k=p+1}^{10}
       \{\frac{1}{2}\langle\partial_{\tilde{a}}X^{i},\partial_{\tilde{a}}X^{i}\rangle +\frac{1}{6}
       \langle F_{\tilde{a}\tilde{b}\tilde{c}},F_{\tilde{a}\tilde{b}\tilde{c}}\rangle+
       \nonumber \\ &&  \frac{1}{2} (\Psi^{ \dag U,i})i\tilde{\gamma}^{\tilde{c}'}\partial_{\tilde{c}'}(\Psi^{ U,i}) +  \frac{1}{6}\Psi^{ \dag U,i} i \tilde{\gamma}^{\tilde{a}\tilde{b}\tilde{c}}\kappa^{i}\partial_{[\tilde{a}}\chi^{ U}_{\tilde{b}\tilde{c}]} 
       \})\Big)^{1/2}.\nonumber \\ && \nonumber \\ &&  E_{extra}\approx 6T_{G0} \int  d^{p+1}x \Big( Tr(  \Sigma_{a,b=0}^{p}\Sigma_{j=p+1}^{10}\Big(\nonumber \\
                &&  
         \{\frac{1}{2}\langle \partial_{b}\partial_{c}\partial_{d}\partial_{e}\partial_{f}X^{i},\partial_{b}\partial_{c}\partial_{d}\partial_{e}\partial_{f}X^{i}\rangle + \frac{1}{2}\sum_{j}(X^{j})^{2}\langle \partial_{c}\partial_{d}\partial_{e}\partial_{f}X^{i},\partial_{c}\partial_{d}\partial_{e}\partial_{f}X^{i}\rangle + \nonumber \\
         &&    \frac{1}{2}\sum_{j}(X^{j})^{4}\langle \partial_{d}\partial_{e}\partial_{f}X^{i},\partial_{d}\partial_{e}\partial_{f}X^{i}\rangle+\frac{1}{2}\sum_{j}(X^{j})^{6}\langle \partial_{e}\partial_{f}X^{i},\partial_{e}\partial_{f}X^{i}\rangle  + \frac{1}{2}\sum_{j}(X^{j})^{8}\langle \partial_{f}X^{i},\partial_{f}X^{i}\rangle + \nonumber \\
         &&\frac{1}{720}
         \langle F^{abca'b'c'},F^{abca'b'c'}\rangle+\frac{1}{120}\sum_{j}(X^{j})^{2}
         \langle F^{abca'b'},F^{abca'b'}\rangle+\frac{1}{24}\sum_{j}(X^{j})^{4}
         \langle F^{abca'},F^{abca'}\rangle + \nonumber \\
         &&\frac{1}{6}\sum_{j}(X^{j})^{6}
         \langle F^{abc},F^{abc}\rangle  +\frac{1}{2}\sum_{j}(X^{j})^{8}
         \langle F^{ab},F^{ab}\rangle -\frac{1}{720}
         \langle[X^{i},X^{j},X^{k},X^{i'},X^{j'},X^{k'}],[X^{i},X^{j},X^{k},X^{i'},X^{j'},X^{k'}]\rangle
         \})\Big)^{1/2}
    \label{sc2}
    \end{eqnarray}
    
 These equation shows that numbers of degrees of freedom on the brane and in the bulk depend on evolutions of scalar, fermions and gauge strings in G-theory. In fact, the interaction of branes causes to evolutions of these fields and change in number of degrees of freedom.   For simplicity, we assume $X^{i}\approx l_{1}$, $\Psi^{i}\approx l_{1}^{3/2}$ and $A^{\mu\nu}\approx l_{2}$ which $l_{1}$ and $l_{2}$ are the separation distance between two branes and the length of branes respectively and only depend on time. Using equations (\ref{sc1}) and (\ref{sc2}), we obtain total energy of system in M-theory:

  \begin{eqnarray}
   && E_{tot,M-theory}=E_{Mp} + E_{anti-Mp} + E_{extra}=\nonumber \\
            && T_{Mp} V\int  dt \Big(\frac{1}{2}(l_{1}')^{2}+\frac{1}{6}(l_{2}')^{2}+\frac{1}{2}l_{1}^{3/2}(l_{1}^{3/2})'+\frac{1}{6}l_{1}^{3/2}(l_{1}^{3/2})'''\Big)^{1/2}+\nonumber \\
                        && T_{Mp} V\int  dt \Big(\frac{1}{2}(l_{1}')^{2}+\frac{1}{6}(l_{2}')^{2}-\frac{1}{2}l_{1}^{3/2}(l_{1}^{3/2})'-\frac{1}{6}l_{1}^{3/2}(l_{1}^{3/2})'''\Big)^{1/2}-\nonumber \\
                                    && 6 T_{G0} V'\int  dt \Big(\frac{l_{1}^{8}}{2}(l_{1}')^{2}+\frac{l_{1}^{8}}{2}(l_{2}')^{2}+
           \frac{l_{1}^{6}}{2}(l_{1}'')^{2}+\frac{l_{1}^{6}}{6}(l_{2}'')^{2}+\nonumber \\
         &&\frac{l_{1}^{4}}{2}(l_{1}''')^{2}+\frac{l_{1}^{4}}{24}(l_{2}''')^{2}+
                    \frac{l_{1}^{2}}{2}(l_{1}'''')^{2}+\frac{l_{1}^{2}}{120}(l_{2}'''')^{2}                       +\nonumber \\
                             &&\frac{1}{2}(l_{1}''''')^{2}+\frac{1}{720}(l_{2}''''')^{2}- \frac{l_{1}^{12}}{720}\Big)^{1/2}.
  \label{sc3}
  \end{eqnarray}
  
  We can obtain the wave equation for the above energy:

    \begin{eqnarray}
     && T_{Mp} V \Big((l_{1}')(l_{1}'')-\frac{3}{4}l_{1}^{1/2}(l_{1}^{3/2})'-\frac{1}{4}l_{1}^{1/2}(l_{1}^{3/2})'''+\frac{3}{4}l_{1}^{2}-\frac{1}{16}\Big)\times\nonumber \\
                               &&  \Big(\frac{1}{2}(l_{1}')^{2}+\frac{1}{6}(l_{2}')^{2}+\frac{1}{2}l_{1}^{3/2}(l_{1}^{3/2})'+\frac{1}{6}l_{1}^{3/2}(l_{1}^{3/2})'''\Big)^{-1/2}+\nonumber \\
                          && T_{Mp} V \Big((l_{1}')(l_{1}'')+\frac{3}{4}l_{1}^{1/2}(l_{1}^{3/2})'+\frac{1}{4}l_{1}^{1/2}(l_{1}^{3/2})'''-\frac{3}{4}l_{1}^{2}+\frac{1}{16}\Big)\times\nonumber \\
                                                         &&  \Big(\frac{1}{2}(l_{1}')^{2}+\frac{1}{6}(l_{2}')^{2}+\frac{1}{2}l_{1}^{3/2}(l_{1}^{3/2})'+\frac{1}{6}l_{1}^{3/2}(l_{1}^{3/2})'''\Big)^{-1/2}-\nonumber \\
                                      && 6 T_{G0} V'\Big(-4l_{1}^{7}(l_{1}')^{2}-4l_{1}^{7}(l_{2}')^{2}-
                                                  3l_{1}^{5}(l_{1}'')^{2}-3l_{1}^{5}(l_{2}'')^{2}-\nonumber \\
                                                 && 2l_{1}^{3}(l_{1}''')^{2}-\frac{l_{1}^{3}}{6}(l_{2}''')^{2}-
                                                            l_{1}(l_{1}'''')^{2}-\frac{l_{1}}{60}(l_{2}'''')^{2}                       -\nonumber \\
                                                                     &&\frac{1}{2}(l_{1}''''')^{2}-\frac{1}{720}(l_{2}''''')^{2}+ \frac{l_{1}^{11}}{360}+\nonumber \\
                                                                                                                                                 &&l_{1}^{8}(l_{1}')-
                                                                                                                                                              l_{1}^{6}(l_{1}'')+l_{1}^{4}(l_{1}''')-
                                                                                                                                                                       l_{1}^{2}(l_{1}'''')                       \Big)                               \times\nonumber \\
                                                                            && \Big(\frac{l_{1}^{8}}{2}(l_{1}')^{2}+\frac{l_{1}^{8}}{2}(l_{2}')^{2}+
             \frac{l_{1}^{6}}{2}(l_{1}'')^{2}+\frac{l_{1}^{6}}{6}(l_{2}'')^{2}+\nonumber \\
           &&\frac{l_{1}^{4}}{2}(l_{1}''')^{2}+\frac{l_{1}^{4}}{24}(l_{2}''')^{2}+
                      \frac{l_{1}^{2}}{2}(l_{1}'''')^{2}+\frac{l_{1}^{2}}{120}(l_{2}'''')^{2}                       +\nonumber \\
                               &&\frac{1}{2}(l_{1}''''')^{2}+\frac{1}{720}(l_{2}''''')^{2}- \frac{l_{1}^{12}}{720}\Big)^{-1/2}=0.
    \label{sc4}
    \end{eqnarray}

     \begin{eqnarray}
      &&  T_{Mp} V \Big(\frac{1}{3}(l_{2}')\Big) \Big(\frac{1}{2}(l_{1}')^{2}+\frac{1}{6}(l_{2}')^{2}+\frac{1}{2}l_{1}^{3/2}(l_{1}^{3/2})'+\frac{1}{6}l_{1}^{3/2}(l_{1}^{3/2})'''\Big)^{-1/2}+\nonumber \\
                           && T_{Mp} V \Big(\frac{1}{3}(l_{2}')\Big) \Big(\frac{1}{2}(l_{1}')^{2}+\frac{1}{6}(l_{2}')^{2}-\frac{1}{2}l_{1}^{3/2}(l_{1}^{3/2})'-\frac{1}{6}l_{1}^{3/2}(l_{1}^{3/2})'''\Big)^{1/2}-\nonumber \\
                                       && 6 T_{G0} V'\Big(l_{1}^{8}(l_{2}')-\frac{l_{1}^{6}}{3}(l_{2}'')+\frac{l_{1}^{4}}{12}(l_{2}''')-\frac{l_{1}^{2}}{60}(l_{2}'''')                       +\frac{1}{360}(l_{2}''''')\Big)\times \nonumber \\
                                                   && \Big(\frac{l_{1}^{8}}{2}(l_{1}')^{2}+\frac{l_{1}^{8}}{2}(l_{2}')^{2}+
              \frac{l_{1}^{6}}{2}(l_{1}'')^{2}+\frac{l_{1}^{6}}{6}(l_{2}'')^{2}+\nonumber \\
            &&\frac{l_{1}^{4}}{2}(l_{1}''')^{2}+\frac{l_{1}^{4}}{24}(l_{2}''')^{2}+
                       \frac{l_{1}^{2}}{2}(l_{1}'''')^{2}+\frac{l_{1}^{2}}{120}(l_{2}'''')^{2}                       +\nonumber \\
                                &&\frac{1}{2}(l_{1}''''')^{2}+\frac{1}{720}(l_{2}''''')^{2}- \frac{l_{1}^{12}}{720}\Big)^{-1/2}=0.
     \label{sc5}
     \end{eqnarray} 
     
     The approximate solutions of above equations are:

       \begin{eqnarray}
        && l_{1}\approx e^{-\frac{2T_{Mp}V t}{3T_{G0}V'(t_{s}-t)}}[(t^{-2/3}-t_{s}^{-2/3})^{6}+ 6(t^{-2/3}-t_{s}^{-2/3})^{4}+\nonumber \\ && 120(t^{-2/3}-t_{s}^{-2/3})^{2}+720(t^{-2/3}-t_{s}^{-2/3})]
          \nonumber \\ &&\nonumber \\ &&  l_{2}\approx t^{6}e^{\frac{T_{Mp}Vt}{T_{G0}V'(t_{s}-t)}}(1+\frac{T_{Mp}V t}{T_{G0}V'(t_{s}-t)}l_{1}^{-2}+(\frac{T_{Mp}V t}{T_{G0}V'(t_{s}-t)})^{2}l_{1}^{-4}+\nonumber \\ &&(\frac{T_{Mp}V t}{T_{G0}V'(t_{s}-t)})^{3}l_{1}^{-6}+(\frac{T_{Mp}V t}{T_{G0}V'(t_{s}-t)})^{4}l_{1}^{-8})
       \label{sc6}
       \end{eqnarray}
       
where $t_{s}$ is time of collision of two branes. It is clear that by passing time, the length of branes ($l_{2}$) increases and the separation distance between branes ($l_{1}$) decreases and shrinks to zero at colliding time ($t=t_{s}$). This is because that at $t=0$, there is nothing. Then, two strings with negative and positive energies are produced. These objects interact with each other, excited and produce extra degrees of freedom in another dimensions. Coincidence with the emergence of degrees of freedom, Gp-branes are produced which are reduced to Mp-branes by compactification and some extra energies are created. By dissolving extra energy in Mp-branes, the length of these branes grows and tends to infinity at the point of colliding of branes. Now, by substituting equation (\ref{sc6}) in equation (\ref{sc1}) we can calculate number of degrees of freedom on the surface:

  \begin{eqnarray}
   && N_{sur}\approx E_{Mp} \approx T_{Mp} \Big(\frac{T_{Mp}V}{T_{G0}V'(t_{s}-t)}[t^{6}e^{\frac{T_{Mp}Vt}{T_{G0}V'(t_{s}-t)}}(1+\frac{T_{Mp}V t}{T_{G0}V'(t_{s}-t)}l_{1}^{-2}+(\frac{T_{Mp}V t}{T_{G0}V'(t_{s}-t)})^{2}l_{1}^{-4}+\nonumber \\ &&(\frac{T_{Mp}V t}{T_{G0}V'(t_{s}-t)})^{3}l_{1}^{-6}+(\frac{T_{Mp}V t}{T_{G0}V'(t_{s}-t)})^{4}l_{1}^{-8})]+\nonumber \\ && [6t^{5}e^{\frac{T_{Mp}Vt}{T_{G0}V'(t_{s}-t)}}(1+\frac{T_{Mp}V t}{T_{G0}V'(t_{s}-t)}l_{1}^{-2}+(\frac{T_{Mp}V t}{T_{G0}V'(t_{s}-t)})^{2}l_{1}^{-4}+\nonumber \\ &&(\frac{T_{Mp}V t}{T_{G0}V'(t_{s}-t)})^{3}l_{1}^{-6}+(\frac{T_{Mp}V t}{T_{G0}V'(t_{s}-t)})^{4}l_{1}^{-8})]+\nonumber \\ && [t^{6}e^{\frac{T_{Mp}Vt}{T_{G0}V'(t_{s}-t)}}l_{1}'(\frac{T_{Mp}V t}{T_{G0}V'(t_{s}-t)}l_{1}^{-3}+(\frac{T_{Mp}V t}{T_{G0}V'(t_{s}-t)})^{2}l_{1}^{-5}+\nonumber \\ &&(\frac{T_{Mp}V t}{T_{G0}V'(t_{s}-t)})^{3}l_{1}^{-7}+(\frac{T_{Mp}V t}{T_{G0}V'(t_{s}-t)})^{4}l_{1}^{-9})]+\nonumber \\ &&
   [t^{6}e^{\frac{T_{Mp}Vt}{T_{G0}V'(t_{s}-t)}}(\frac{T_{Mp}V }{T_{G0}V'(t_{s}-t)}l_{1}^{-2}+(\frac{T_{Mp}V }{T_{G0}V'(t_{s}-t)})^{2}l_{1}^{-4}+\nonumber \\ &&(\frac{T_{Mp}V }{T_{G0}V'(t_{s}-t)})^{3}l_{1}^{-6}+(\frac{T_{Mp}V }{T_{G0}V'(t_{s}-t)})^{4}l_{1}^{-8})]+\nonumber \\ && 
   [t^{6}e^{\frac{T_{Mp}Vt}{T_{G0}V'(t_{s}-t)}}(\frac{T_{Mp}V t}{T_{G0}V'(t_{s}-t)^{2}}l_{1}^{-2}+(\frac{T_{Mp}V t}{T_{G0}V'(t_{s}-t)^{3/2}})^{2}l_{1}^{-4}+\nonumber \\ &&(\frac{T_{Mp}V t}{T_{G0}V'(t_{s}-t)^{4/3}})^{3}l_{1}^{-6}+(\frac{T_{Mp}V t}{T_{G0}V'(t_{s}-t)^{5/4}})^{4}l_{1}^{-8})]\Big)\times\nonumber \\ && 
   \Big(t^{6}e^{\frac{T_{Mp}Vt}{T_{G0}V'(t_{s}-t)}}(1+\frac{T_{Mp}V t}{T_{G0}V'(t_{s}-t)}l_{1}^{-2}+(\frac{T_{Mp}V t}{T_{G0}V'(t_{s}-t)})^{2}l_{1}^{-4}+\nonumber \\ &&(\frac{T_{Mp}V t}{T_{G0}V'(t_{s}-t)})^{3}l_{1}^{-6}+(\frac{T_{Mp}V t}{T_{G0}V'(t_{s}-t)})^{4}l_{1}^{-8})-\frac{3}{4}l_{1}^{2}l_{1}'+l_{1}l_{1}'\Big).
  \label{sc7}
  \end{eqnarray}
  
  This equation shows that the energy  of brane and number of degrees of freedom  is zero at $t=0$, grows by time and tends to infinity at time of collision of branes ($t=t_{s}$). In fact, it is possible that two types of branes with different physics, one with a time coordinate and another one with two time coordinates are produced from nothing in fourteen dimensions such as the sum over energy of them be zero and then these objects are compactified on three circle and Mp-branes are created. After that, universe is born on one of these branes and expands. To show this, we put p=3 and use of following equations:

         \begin{eqnarray}
          && N_{sur}=\frac{4\pi r_{H}^{2}}{l_{p}^{2}} \quad r_{H}\approx \frac{1}{H} \quad H=\frac{\dot{a}}{a}\nonumber \\ && \Rightarrow a=a_{0}e^{-\int dt (\frac{4\pi}{l_{p}^{2}N_{sur}})^{1/2}} =a_{0}e^{-\int dt (\frac{4\pi}{l_{p}^{2}E_{M3}})^{1/2}}=a_{0}e^{-(\frac{4\pi}{l_{p}^{2}})^{1/2}F}\nonumber \\ && F\approx \Big(t^{6}e^{\frac{T_{Mp}Vt}{T_{G0}V'(t_{s}-t)}}(1+\frac{T_{Mp}V t}{T_{G0}V'(t_{s}-t)}l_{1}^{-2}+(\frac{T_{Mp}V t}{T_{G0}V'(t_{s}-t)})^{2}l_{1}^{-4}+\nonumber \\ &&(\frac{T_{Mp}V t}{T_{G0}V'(t_{s}-t)})^{3}l_{1}^{-6}+(\frac{T_{Mp}V t}{T_{G0}V'(t_{s}-t)})^{4}l_{1}^{-8})\Big)^{-1}-\nonumber \\ &&\frac{3}{4}\Big(e^{-\frac{2T_{Mp}V t}{3T_{G0}V'(t_{s}-t)}}[(t^{-2/3}-t_{s}^{-2/3})^{6}+ 6(t^{-2/3}-t_{s}^{-2/3})^{4}+\nonumber \\ && 120(t^{-2/3}-t_{s}^{-2/3})^{2}+720(t^{-2/3}-t_{s}^{-2/3})]\Big)^{-3}
         \label{sc8}
         \end{eqnarray}

   where $H$  is the Hubble Parameter, $a$  is the scale factor of universe and $r_{H}$  is the event horizon of universe. This equation shows that coincidence with the birth of Mp-branes from compactifying Gp-branes, universe is produced, grows and achieve to the present size. In fact, the scale factor of universe depends on the parameters of G-model and M-theory and evolves from nothing to expansion phase in four stages: In first stage, two one dimensional objects like two strings are created in 14th dimension. Second, these strings are excited in other 13 dimensions and construct Gp-branes that their dimensions can change from zero to fourteen. Third, these Gp-branes are compactified on three circles and Mp-branes are created. Fourth, universe is born on one of Mp-branes and expands as due to inequality between number of degrees of freedom on the universe surface and in the bulk in Padmanabhan model.

  \section{Testing the model with observations }\label{o3}
  
 Recent experiments \cite{kj1} predict that the magnitude of the slow-roll parameters and the tensor-to-scalar ratio should be very smaller than one. In this section, we will show that G-theory gives the correct values for cosmological parameters during inflation era. Using definitions of \cite{kj2} and equation (\ref{sc8}), we can obtain the slow-roll parameters and the tensor-to-scalar ratio and compare with previous predictions:

          \begin{eqnarray}
         && H=\frac{\dot{a}}{a}=-(\frac{4\pi}{l_{p}^{2}})^{1/2}\dot{F} \Rightarrow \nonumber \\  && \varepsilon=-\frac{\dot{H}}{H^{2}}=(\frac{4\pi}{l_{p}^{2}})^{-1/2}\frac{\ddot{F}}{(\dot{F})^{2}} \nonumber \\ &&=(\frac{4\pi}{l_{p}^{2}})^{-1/2}[[\Big(20t^{4}e^{\frac{T_{Mp}Vt}{T_{G0}V'(t_{s}-t)}}(1+\frac{T_{Mp}V t}{T_{G0}V'(t_{s}-t)}l_{1}^{-2}+(\frac{T_{Mp}V t}{T_{G0}V'(t_{s}-t)})^{2}l_{1}^{-4}+\nonumber \\ &&(\frac{T_{Mp}V t}{T_{G0}V'(t_{s}-t)})^{3}l_{1}^{-6}+(\frac{T_{Mp}V t}{T_{G0}V'(t_{s}-t)})^{4}l_{1}^{-8})\Big)+\nonumber \\ && \Big(t^{6}e^{\frac{T_{Mp}Vt}{T_{G0}V'(t_{s}-t)}}(\frac{T_{Mp}V t}{T_{G0}V'(t_{s}-t)^{3}}l_{1}^{-2}+2(\frac{T_{Mp}V t}{T_{G0}V'(t_{s}-t)})^{2}(t_{s}-t)^{-2}l_{1}^{-4}+\nonumber \\ &&3(\frac{T_{Mp}V t}{T_{G0}V'(t_{s}-t)})^{3}(t_{s}-t)^{-2}l_{1}^{-6}+4(\frac{T_{Mp}V t}{T_{G0}V'(t_{s}-t)})^{4}(t_{s}-t)^{-2}l_{1}^{-8})\Big)]\times\nonumber \\ &&\Big(t^{6}e^{\frac{T_{Mp}Vt}{T_{G0}V'(t_{s}-t)}}(1+\frac{T_{Mp}V t}{T_{G0}V'(t_{s}-t)}l_{1}^{-2}+(\frac{T_{Mp}V t}{T_{G0}V'(t_{s}-t)})^{2}l_{1}^{-4}+\nonumber \\ &&(\frac{T_{Mp}V t}{T_{G0}V'(t_{s}-t)})^{3}l_{1}^{-6}+(\frac{T_{Mp}V t}{T_{G0}V'(t_{s}-t)})^{4}l_{1}^{-8})\Big)^{-3}-\nonumber \\ &&\frac{3}{4}\Big(e^{-\frac{2T_{Mp}V t}{3T_{G0}V'(t_{s}-t)}}[\frac{2}{3}t^{-7/3}(t^{-2/3}-t_{s}^{-2/3})^{5}+ 4t^{-7/3}(t^{-2/3}-t_{s}^{-2/3})^{3}+\nonumber \\ && 80t^{-7/3}(t^{-2/3}-t_{s}^{-2/3})+\frac{5760}{9}t^{-7/3}]\Big)\times\nonumber \\ &&\Big(e^{-\frac{2T_{Mp}V t}{3T_{G0}V'(t_{s}-t)}}[(t^{-2/3}-t_{s}^{-2/3})^{6}+ 6(t^{-2/3}-t_{s}^{-2/3})^{4}+\nonumber \\ && 120(t^{-2/3}-t_{s}^{-2/3})^{2}+720(t^{-2/3}-t_{s}^{-2/3})]\Big)^{-4}]\times\nonumber \\ &&[[\Big(5t^{5}e^{\frac{T_{Mp}Vt}{T_{G0}V'(t_{s}-t)}}(1+\frac{T_{Mp}V t}{T_{G0}V'(t_{s}-t)}l_{1}^{-2}+(\frac{T_{Mp}V t}{T_{G0}V'(t_{s}-t)})^{2}l_{1}^{-4}+\nonumber \\ &&(\frac{T_{Mp}V t}{T_{G0}V'(t_{s}-t)})^{3}l_{1}^{-6}+(\frac{T_{Mp}V t}{T_{G0}V'(t_{s}-t)})^{4}l_{1}^{-8})\Big)+\nonumber \\ && \Big(t^{6}e^{\frac{T_{Mp}Vt}{T_{G0}V'(t_{s}-t)}}(\frac{T_{Mp}V t}{T_{G0}V'(t_{s}-t)^{2}}l_{1}^{-2}+2(\frac{T_{Mp}V t}{T_{G0}V'(t_{s}-t)})^{2}(t_{s}-t)^{-1}l_{1}^{-4}+\nonumber \\ &&3(\frac{T_{Mp}V t}{T_{G0}V'(t_{s}-t)})^{3}(t_{s}-t)^{-1}l_{1}^{-6}+4(\frac{T_{Mp}V t}{T_{G0}V'(t_{s}-t)})^{4}(t_{s}-t)^{-1}l_{1}^{-8})\Big)]\times\nonumber \\ &&\Big(t^{6}e^{\frac{T_{Mp}Vt}{T_{G0}V'(t_{s}-t)}}(1+\frac{T_{Mp}V t}{T_{G0}V'(t_{s}-t)}l_{1}^{-2}+(\frac{T_{Mp}V t}{T_{G0}V'(t_{s}-t)})^{2}l_{1}^{-4}+\nonumber \\ &&(\frac{T_{Mp}V t}{T_{G0}V'(t_{s}-t)})^{3}l_{1}^{-6}+(\frac{T_{Mp}V t}{T_{G0}V'(t_{s}-t)})^{4}l_{1}^{-8})\Big)^{-2}-\nonumber \\ &&\frac{3}{4}\Big(e^{-\frac{2T_{Mp}V t}{3T_{G0}V'(t_{s}-t)}}[\frac{2}{3}t^{-4/3}(t^{-2/3}-t_{s}^{-2/3})^{5}+ 4t^{-4/3}(t^{-2/3}-t_{s}^{-2/3})^{3}+\nonumber \\ && 80t^{-4/3}(t^{-2/3}-t_{s}^{-2/3})+\frac{1440}{3}t^{-4/3}(t^{-2/3}-t_{s}^{-2/3})]\Big)\times\nonumber \\ &&\Big(e^{-\frac{2T_{Mp}V t}{3T_{G0}V'(t_{s}-t)}}[(t^{-2/3}-t_{s}^{-2/3})^{6}+ 6(t^{-2/3}-t_{s}^{-2/3})^{4}+\nonumber \\ && 120(t^{-2/3}-t_{s}^{-2/3})^{2}+720(t^{-2/3}-t_{s}^{-2/3})]\Big)^{-3}]^{-2}\nonumber \\ && \nonumber \\ &&\nonumber \\ && \eta=-\frac{\ddot{H}}{2H\dot{H}}=(\frac{4\pi}{l_{p}^{2}})^{-1/2}\frac{\dddot{F}}{2\dot{F}\ddot{F}}=\nonumber \\ && 2^{-1}(\frac{4\pi}{l_{p}^{2}})^{-1/2}[[\Big(20t^{4}e^{\frac{T_{Mp}Vt}{T_{G0}V'(t_{s}-t)}}(1+\frac{T_{Mp}V t}{T_{G0}V'(t_{s}-t)}l_{1}^{-2}+(\frac{T_{Mp}V t}{T_{G0}V'(t_{s}-t)})^{2}l_{1}^{-4}+\nonumber \\ &&(\frac{T_{Mp}V t}{T_{G0}V'(t_{s}-t)})^{3}l_{1}^{-6}+(\frac{T_{Mp}V t}{T_{G0}V'(t_{s}-t)})^{4}l_{1}^{-8})\Big)+\nonumber \\ && \Big(t^{6}e^{\frac{T_{Mp}Vt}{T_{G0}V'(t_{s}-t)}}(\frac{T_{Mp}V t}{T_{G0}V'(t_{s}-t)^{3}}l_{1}^{-2}+2(\frac{T_{Mp}V t}{T_{G0}V'(t_{s}-t)})^{2}(t_{s}-t)^{-2}l_{1}^{-4}+\nonumber \\ &&3(\frac{T_{Mp}V t}{T_{G0}V'(t_{s}-t)})^{3}(t_{s}-t)^{-2}l_{1}^{-6}+4(\frac{T_{Mp}V t}{T_{G0}V'(t_{s}-t)})^{4}(t_{s}-t)^{-2}l_{1}^{-8})\Big)]\times\nonumber \\ &&\Big(t^{6}e^{\frac{T_{Mp}Vt}{T_{G0}V'(t_{s}-t)}}(1+\frac{T_{Mp}V t}{T_{G0}V'(t_{s}-t)}l_{1}^{-2}+(\frac{T_{Mp}V t}{T_{G0}V'(t_{s}-t)})^{2}l_{1}^{-4}+\nonumber \\ &&(\frac{T_{Mp}V t}{T_{G0}V'(t_{s}-t)})^{3}l_{1}^{-6}+(\frac{T_{Mp}V t}{T_{G0}V'(t_{s}-t)})^{4}l_{1}^{-8})\Big)^{-3}-\nonumber \\ &&\frac{3}{4}\Big(e^{-\frac{2T_{Mp}V t}{3T_{G0}V'(t_{s}-t)}}[\frac{2}{3}t^{-7/3}(t^{-2/3}-t_{s}^{-2/3})^{5}+ 4t^{-7/3}(t^{-2/3}-t_{s}^{-2/3})^{3}+\nonumber \\ && 80t^{-7/3}(t^{-2/3}-t_{s}^{-2/3})+\frac{5760}{9}t^{-7/3}]\Big)\times\nonumber \\ &&\Big(e^{-\frac{2T_{Mp}V t}{3T_{G0}V'(t_{s}-t)}}[(t^{-2/3}-t_{s}^{-2/3})^{6}+ 6(t^{-2/3}-t_{s}^{-2/3})^{4}+\nonumber \\ && 120(t^{-2/3}-t_{s}^{-2/3})^{2}+720(t^{-2/3}-t_{s}^{-2/3})]\Big)^{-4}]^{-1}\times\nonumber \\ &&[[\Big(5t^{5}e^{\frac{T_{Mp}Vt}{T_{G0}V'(t_{s}-t)}}(1+\frac{T_{Mp}V t}{T_{G0}V'(t_{s}-t)}l_{1}^{-2}+(\frac{T_{Mp}V t}{T_{G0}V'(t_{s}-t)})^{2}l_{1}^{-4}+\nonumber \\ &&(\frac{T_{Mp}V t}{T_{G0}V'(t_{s}-t)})^{3}l_{1}^{-6}+(\frac{T_{Mp}V t}{T_{G0}V'(t_{s}-t)})^{4}l_{1}^{-8})\Big)+\nonumber \\ && \Big(t^{6}e^{\frac{T_{Mp}Vt}{T_{G0}V'(t_{s}-t)}}(\frac{T_{Mp}V t}{T_{G0}V'(t_{s}-t)^{2}}l_{1}^{-2}+2(\frac{T_{Mp}V t}{T_{G0}V'(t_{s}-t)})^{2}(t_{s}-t)^{-1}l_{1}^{-4}+\nonumber \\ &&3(\frac{T_{Mp}V t}{T_{G0}V'(t_{s}-t)})^{3}(t_{s}-t)^{-1}l_{1}^{-6}+4(\frac{T_{Mp}V t}{T_{G0}V'(t_{s}-t)})^{4}(t_{s}-t)^{-1}l_{1}^{-8})\Big)]\times\nonumber \\ &&\Big(t^{6}e^{\frac{T_{Mp}Vt}{T_{G0}V'(t_{s}-t)}}(1+\frac{T_{Mp}V t}{T_{G0}V'(t_{s}-t)}l_{1}^{-2}+(\frac{T_{Mp}V t}{T_{G0}V'(t_{s}-t)})^{2}l_{1}^{-4}+\nonumber \\ &&(\frac{T_{Mp}V t}{T_{G0}V'(t_{s}-t)})^{3}l_{1}^{-6}+(\frac{T_{Mp}V t}{T_{G0}V'(t_{s}-t)})^{4}l_{1}^{-8})\Big)^{-2}-\nonumber \\ &&\frac{3}{4}\Big(e^{-\frac{2T_{Mp}V t}{3T_{G0}V'(t_{s}-t)}}[\frac{2}{3}t^{-4/3}(t^{-2/3}-t_{s}^{-2/3})^{5}+ 4t^{-4/3}(t^{-2/3}-t_{s}^{-2/3})^{3}+\nonumber \\ && 80t^{-4/3}(t^{-2/3}-t_{s}^{-2/3})+\frac{1440}{3}t^{-4/3}(t^{-2/3}-t_{s}^{-2/3})]\Big)\times\nonumber \\ &&\Big(e^{-\frac{2T_{Mp}V t}{3T_{G0}V'(t_{s}-t)}}[(t^{-2/3}-t_{s}^{-2/3})^{6}+ 6(t^{-2/3}-t_{s}^{-2/3})^{4}+\nonumber \\ && 120(t^{-2/3}-t_{s}^{-2/3})^{2}+720(t^{-2/3}-t_{s}^{-2/3})]\Big)^{-3}]^{-1}\times\nonumber \\ &&[[\Big(80t^{3}e^{\frac{T_{Mp}Vt}{T_{G0}V'(t_{s}-t)}}(1+\frac{T_{Mp}V t}{T_{G0}V'(t_{s}-t)}l_{1}^{-2}+(\frac{T_{Mp}V t}{T_{G0}V'(t_{s}-t)})^{2}l_{1}^{-4}+\nonumber \\ &&(\frac{T_{Mp}V t}{T_{G0}V'(t_{s}-t)})^{3}l_{1}^{-6}+(\frac{T_{Mp}V t}{T_{G0}V'(t_{s}-t)})^{4}l_{1}^{-8})\Big)+\nonumber \\ && \Big(t^{6}e^{\frac{T_{Mp}Vt}{T_{G0}V'(t_{s}-t)}}(\frac{T_{Mp}V t}{T_{G0}V'(t_{s}-t)^{4}}l_{1}^{-2}+2(\frac{T_{Mp}V t}{T_{G0}V'(t_{s}-t)})^{2}(t_{s}-t)^{-3}l_{1}^{-4}+\nonumber \\ &&3(\frac{T_{Mp}V t}{T_{G0}V'(t_{s}-t)})^{3}(t_{s}-t)^{-3}l_{1}^{-6}+4(\frac{T_{Mp}V t}{T_{G0}V'(t_{s}-t)})^{4}(t_{s}-t)^{-3}l_{1}^{-8})\Big)]\times\nonumber \\ &&\Big(t^{6}e^{\frac{T_{Mp}Vt}{T_{G0}V'(t_{s}-t)}}(1+\frac{T_{Mp}V t}{T_{G0}V'(t_{s}-t)}l_{1}^{-2}+(\frac{T_{Mp}V t}{T_{G0}V'(t_{s}-t)})^{2}l_{1}^{-4}+\nonumber \\ &&(\frac{T_{Mp}V t}{T_{G0}V'(t_{s}-t)})^{3}l_{1}^{-6}+(\frac{T_{Mp}V t}{T_{G0}V'(t_{s}-t)})^{4}l_{1}^{-8})\Big)^{-4}-\nonumber \\ &&\frac{3}{4}\Big(e^{-\frac{2T_{Mp}V t}{3T_{G0}V'(t_{s}-t)}}[\frac{2}{3}t^{-10/3}(t^{-2/3}-t_{s}^{-2/3})^{5}+ 4t^{-10/3}(t^{-2/3}-t_{s}^{-2/3})^{3}+\nonumber \\ && 80t^{-10/3}(t^{-2/3}-t_{s}^{-2/3})+\frac{17180}{27}t^{-10/3}]\Big)\times\nonumber \\ &&\Big(e^{-\frac{2T_{Mp}V t}{3T_{G0}V'(t_{s}-t)}}[(t^{-2/3}-t_{s}^{-2/3})^{6}+ 6(t^{-2/3}-t_{s}^{-2/3})^{4}+\nonumber \\ && 120(t^{-2/3}-t_{s}^{-2/3})^{2}+720(t^{-2/3}-t_{s}^{-2/3})]\Big)^{-5}] \nonumber \\ && \nonumber \\ &&\nonumber \\ && r= 16\varepsilon=\nonumber \\ &&16(\frac{4\pi}{l_{p}^{2}})^{-1/2}[[\Big(20t^{4}e^{\frac{T_{Mp}Vt}{T_{G0}V'(t_{s}-t)}}(1+\frac{T_{Mp}V t}{T_{G0}V'(t_{s}-t)}l_{1}^{-2}+(\frac{T_{Mp}V t}{T_{G0}V'(t_{s}-t)})^{2}l_{1}^{-4}+\nonumber \\ &&(\frac{T_{Mp}V t}{T_{G0}V'(t_{s}-t)})^{3}l_{1}^{-6}+(\frac{T_{Mp}V t}{T_{G0}V'(t_{s}-t)})^{4}l_{1}^{-8})\Big)+\nonumber \\ && \Big(t^{6}e^{\frac{T_{Mp}Vt}{T_{G0}V'(t_{s}-t)}}(\frac{T_{Mp}V t}{T_{G0}V'(t_{s}-t)^{3}}l_{1}^{-2}+2(\frac{T_{Mp}V t}{T_{G0}V'(t_{s}-t)})^{2}(t_{s}-t)^{-2}l_{1}^{-4}+\nonumber \\ &&3(\frac{T_{Mp}V t}{T_{G0}V'(t_{s}-t)})^{3}(t_{s}-t)^{-2}l_{1}^{-6}+4(\frac{T_{Mp}V t}{T_{G0}V'(t_{s}-t)})^{4}(t_{s}-t)^{-2}l_{1}^{-8})\Big)]\times\nonumber \\ &&\Big(t^{6}e^{\frac{T_{Mp}Vt}{T_{G0}V'(t_{s}-t)}}(1+\frac{T_{Mp}V t}{T_{G0}V'(t_{s}-t)}l_{1}^{-2}+(\frac{T_{Mp}V t}{T_{G0}V'(t_{s}-t)})^{2}l_{1}^{-4}+\nonumber \\ &&(\frac{T_{Mp}V t}{T_{G0}V'(t_{s}-t)})^{3}l_{1}^{-6}+(\frac{T_{Mp}V t}{T_{G0}V'(t_{s}-t)})^{4}l_{1}^{-8})\Big)^{-3}-\nonumber \\ &&\frac{3}{4}\Big(e^{-\frac{2T_{Mp}V t}{3T_{G0}V'(t_{s}-t)}}[\frac{2}{3}t^{-7/3}(t^{-2/3}-t_{s}^{-2/3})^{5}+ 4t^{-7/3}(t^{-2/3}-t_{s}^{-2/3})^{3}+\nonumber \\ && 80t^{-7/3}(t^{-2/3}-t_{s}^{-2/3})+\frac{5760}{9}t^{-7/3}]\Big)\times\nonumber \\ &&\Big(e^{-\frac{2T_{Mp}V t}{3T_{G0}V'(t_{s}-t)}}[(t^{-2/3}-t_{s}^{-2/3})^{6}+ 6(t^{-2/3}-t_{s}^{-2/3})^{4}+\nonumber \\ && 120(t^{-2/3}-t_{s}^{-2/3})^{2}+720(t^{-2/3}-t_{s}^{-2/3})]\Big)^{-4}]\times\nonumber \\ &&[[\Big(5t^{5}e^{\frac{T_{Mp}Vt}{T_{G0}V'(t_{s}-t)}}(1+\frac{T_{Mp}V t}{T_{G0}V'(t_{s}-t)}l_{1}^{-2}+(\frac{T_{Mp}V t}{T_{G0}V'(t_{s}-t)})^{2}l_{1}^{-4}+\nonumber \\ &&(\frac{T_{Mp}V t}{T_{G0}V'(t_{s}-t)})^{3}l_{1}^{-6}+(\frac{T_{Mp}V t}{T_{G0}V'(t_{s}-t)})^{4}l_{1}^{-8})\Big)+\nonumber \\ && \Big(t^{6}e^{\frac{T_{Mp}Vt}{T_{G0}V'(t_{s}-t)}}(\frac{T_{Mp}V t}{T_{G0}V'(t_{s}-t)^{2}}l_{1}^{-2}+2(\frac{T_{Mp}V t}{T_{G0}V'(t_{s}-t)})^{2}(t_{s}-t)^{-1}l_{1}^{-4}+\nonumber \\ &&3(\frac{T_{Mp}V t}{T_{G0}V'(t_{s}-t)})^{3}(t_{s}-t)^{-1}l_{1}^{-6}+4(\frac{T_{Mp}V t}{T_{G0}V'(t_{s}-t)})^{4}(t_{s}-t)^{-1}l_{1}^{-8})\Big)]\times\nonumber \\ &&\Big(t^{6}e^{\frac{T_{Mp}Vt}{T_{G0}V'(t_{s}-t)}}(1+\frac{T_{Mp}V t}{T_{G0}V'(t_{s}-t)}l_{1}^{-2}+(\frac{T_{Mp}V t}{T_{G0}V'(t_{s}-t)})^{2}l_{1}^{-4}+\nonumber \\ &&(\frac{T_{Mp}V t}{T_{G0}V'(t_{s}-t)})^{3}l_{1}^{-6}+(\frac{T_{Mp}V t}{T_{G0}V'(t_{s}-t)})^{4}l_{1}^{-8})\Big)^{-2}-\nonumber \\ &&\frac{3}{4}\Big(e^{-\frac{2T_{Mp}V t}{3T_{G0}V'(t_{s}-t)}}[\frac{2}{3}t^{-4/3}(t^{-2/3}-t_{s}^{-2/3})^{5}+ 4t^{-4/3}(t^{-2/3}-t_{s}^{-2/3})^{3}+\nonumber \\ && 80t^{-4/3}(t^{-2/3}-t_{s}^{-2/3})+\frac{1440}{3}t^{-4/3}(t^{-2/3}-t_{s}^{-2/3})]\Big)\times\nonumber \\ &&\Big(e^{-\frac{2T_{Mp}V t}{3T_{G0}V'(t_{s}-t)}}[(t^{-2/3}-t_{s}^{-2/3})^{6}+ 6(t^{-2/3}-t_{s}^{-2/3})^{4}+\nonumber \\ && 120(t^{-2/3}-t_{s}^{-2/3})^{2}+720(t^{-2/3}-t_{s}^{-2/3})]\Big)^{-3}]^{-2}
          \label{kj1}
          \end{eqnarray}
          
   It is clear that during inflation era, the age of universe (t) is very smaller respect to time of collision between branes ($t_{s}$) and consequently, ($(t_{s}-t)^{-m}\ll 1$) which m is an integer number. Using equation (\ref{kj2}), we can obtain  the following results:

             \begin{eqnarray}
              && 0\ll t\ll t_{s}\Rightarrow   \eta \approx \Sigma B_{n}t^{-n} (t_{s}-t)^{-n}\ll 1 \nonumber\\&& \varepsilon\approx \Sigma A_{m}t^{-m} (t_{s}-t)^{-m}\ll 1\Rightarrow  r\ll 1\label{kj2}
             \end{eqnarray}
                
 These results show that   the magnitude of the slow-roll parameters and the tensor-to-scalar ratio are very smaller than one which is in agreement with predictions of experiments in ref.\cite{kj1,kj1t1,kj1t2}. Thus, G-theory produces the correct values for cosmological parameters and can explain the phenomenological events.
 
   \section{Extended theories of gravity without anomaly in G-theory }\label{o4}
   
   Previously, some effective Dark energy models have been proposed that were in very good agreement with observations \cite{kj3,kj4}. In this section, we consider the origin of these models in G-theory and show that they may be anomaly-free. To this end, we will begin with eleven dimensional manifold in Horava-Witten mechanism and add one three dimensional manifold related to Lie-three-algebra. We will assert that some CGG terms are produced on this new 14-dimensional manifold that cancel the anomalies in eleven-dimensional supergravity and produce the extended theories of gravity.

  First, we introduce the Horava-Witten mechanism in  eleven dimensional space-time. In this model, the bosonic part of the action in 11-dimensional supergravity is given by \cite{b1,b2}:
         
     \begin{eqnarray}
     && S_{Bosonic-SUGRA} = \frac{1}{\bar{\kappa}^{2}}\int d^{11}x\sqrt{g}\Big(-\frac{1}{2}R-\frac{1}{48}G_{IJKL}G^{IJKL}\Big) + S_{CKK} \nonumber\\&& S_{CGG}=-\frac{\sqrt{2}}{3456\bar{\kappa}^{2}}\int_{M^{11}}d^{11}x \varepsilon^{I_{1}I_{2}...I_{11}}C_{I_{1}I_{2}I_{3}}G_{I_{4}...I_{7}}G_{I_{8}...I_{11}}  \label{s1}
     \end{eqnarray} 
     
     where,  $G_{IJKL}$ and $C_{I_{1}I_{2}I_{3}}$  have a direct relation with  the  gauge field $A$, field strength $F$ and the curvature ($R$) \cite{b2}:
     
     \begin{eqnarray}
      && G_{IJKL}=-\frac{3}{\sqrt{2}}\frac{\kappa^{2}}{\lambda^{2}}\varepsilon(x^{11})(F_{IJ}F_{KL}-R_{IJ}R_{KL})+... \nonumber\\&& \delta C_{ABC} =-\frac{\kappa^{2}}{6\sqrt{2}\lambda^{2}}\delta (x^{11}) tr( \epsilon F_{AB}-\epsilon R_{AB})\nonumber\\&& G_{11ABC}=\partial_{11}C_{ABC}+....\nonumber\\&& F^{IJ}=\partial_{I}A^{J}-\partial_{J}A^{I}=\epsilon^{IJ}\partial_{I}A^{J} \nonumber\\&& R_{IJ}=\partial_{I}\Gamma^{\beta}_{J\beta}-\partial_{J}\Gamma^{\beta}_{I\beta} +\Gamma^{\alpha}_{J\beta}\Gamma^{\beta}_{I\alpha} -\Gamma^{\alpha}_{I\beta}\Gamma^{\beta}_{J\alpha}\nonumber\\&& \Gamma_{IJK}=\partial_{I}g_{JK}+\partial_{K}g_{IJ}-\partial_{J}g_{IK}\label{s2}
      \end{eqnarray} 
      
     Here,  $\varepsilon(x^{11})$ is 1 for $x^{11}> 0$ and −1 for $x^{11}< 0$ and also $\delta(x^{11})=\frac{\partial \varepsilon}{\partial x^{11}}$. The gauge variation of the CGG-action, yields the following equation \cite{b2}:
      
      \begin{eqnarray}
        && \delta S_{CGG}|_{11}=-\frac{\sqrt{2}}{3456\bar{\kappa}^{2}}\int_{M^{11}}d^{11}x \varepsilon^{I_{1}I_{2}...I_{11}}\delta C_{I_{1}I_{2}I_{3}}G_{I_{4}...I_{7}}G_{I_{8}...I_{11}}\approx \nonumber\\&& - \frac{\bar{\kappa}^{4}}{128 \lambda^{6}}\int_{M^{10}}\Sigma_{n=1}^{5}(tr F^{n}-tr R^{n}+trF^{n}R^{5-n})\label{s3}
        \end{eqnarray} 
       
      Above terms cancel the  anomaly of  ($S_{Bosonic-SUGRA}$) in eleven dimensional manifold \cite{b2}:
      
      \begin{eqnarray}
          && \delta S_{CGG}|_{11}=-\delta S_{Bosonic-SUGRA}=-S^{anomaly}_{Bosonic-SUGRA}\label{ss3}
          \end{eqnarray}

      Thus, to obtain  the  anomaly-free supergravity in eleven dimensions, we have to use of CGG terms. Now, we answer  this issue that what is the origin of CGG terms in 11-dimensional supergravity. In fact, we propose a theory that CGG terms are appeared in the action of supergravity without adding any by hand. To this end, we  choose a unified shape for all fields by using  Nambu-Poisson brackets and properties of string fields ($X$). We define \cite{q21,A1}:

      \begin{eqnarray}
       &&  X^{I_{i}}=y^{I_{i}}+A^{I_{i}} + \epsilon^{I_{i}}\phi -\epsilon^{I_{i}J}\Gamma^{\alpha}_{\alpha J}\nonumber\\&& \{ X^{I_{i}},X^{I_{j}} \}=\Sigma_{I_{i},I_{j}}\epsilon^{I_{i}I_{j}}\frac{\partial X^{I_{i}}}{\partial y^{I_{j}}}\frac{\partial X^{I_{j}}}{\partial y^{I_{j}}}= \nonumber\\&&\Sigma_{I_{i},I_{j}}\epsilon^{I_{i}I_{j}}\Big(\partial_{I_{i}}A^{I_{j}}-\partial_{I_{i}}(\epsilon^{I_{j}I_{k}}\Gamma^{\alpha}_{\alpha I_{k}})\Big)=F^{I_{i}I_{j}}- R^{I_{i}I_{j}}+ \partial_{I_{j}}\phi\partial_{I_{k}}\phi..  \label{s4}
       \end{eqnarray} 
     
      where $\phi$ is the scalar field, $A$ is the gauge field and $\Gamma$ has the relation with the curvature (R).  Using four-dimensional brackets instead of two-dimensional one, we can obtain the  shape of GG-terms in supergravity in terms of scalar strings ($X$):
              
          \begin{eqnarray}
          && G_{IJKL}= \{ X^{I},X^{J},X^{K},X^{L} \} =\Sigma_{IJKL}\epsilon^{IJKL}\frac{\partial X^{I}}{\partial y^{I}}\frac{\partial X^{J}}{\partial y^{J}}\frac{\partial X^{K}}{\partial y^{K}}\frac{\partial X^{L}}{\partial y^{L}}\nonumber\\&&\Rightarrow \int d^{11}x\sqrt{g}\Big(G_{IJKL}G^{IJKL}\Big)=\nonumber\\&& \int d^{11}x\sqrt{g}\Big(\Sigma_{IJKL}\epsilon^{IJKL}\frac{\partial X^{I}}{\partial y^{I}}\frac{\partial X^{J}}{\partial y^{J}}\frac{\partial X^{K}}{\partial y^{K}}\frac{\partial X^{L}}{\partial y^{L}}\Sigma_{IJKL}\epsilon^{IJKL}\frac{\partial X^{I}}{\partial y^{I}}\frac{\partial X^{J}}{\partial y^{J}}\frac{\partial X^{K}}{\partial y^{K}}\frac{\partial X^{L}}{\partial y^{L}}\Big)  \label{s13}
          \end{eqnarray} 
            
       Above equation help us to extract the CGG terms from GG-terms in supergravity. To this end, we will add a three dimensional manifold  related to Lie-three-algebra to the eleven dimensional supergravity by using the properties of scalar strings ($X$) in Nambu-Poisson brackets \cite{A1}:

           \begin{eqnarray}
           &&X^{I_{i}}=y^{I_{i}}+A^{I_{i}} + \epsilon^{I_{i}}\phi -\epsilon^{I_{i}J}\Gamma^{\alpha}_{\alpha J}\Rightarrow\nonumber\\&& \frac{\partial X^{I_{5}}}{\partial y^{I_{5}}}\approx\delta ( y^{I_{5}})+.. \quad \frac{\partial X^{I_{6}}}{\partial y^{I_{6}}}\approx\delta ( y^{I_{6}})+.. \quad \frac{\partial X^{I_{7}}}{\partial y^{I_{7}}}\approx\delta ( y^{I_{7}})+...\nonumber\\&& \nonumber\\&&\int_{M^{N=3}}\rightarrow\int_{y^{I_{5}}+y^{I_{6}}+y^{I_{7}}}\frac{\partial X^{I_{5}}}{\partial y^{I_{5}}}\frac{\partial X^{I_{6}}}{\partial y^{I_{6}}}\frac{\partial X^{I_{7}}}{\partial y^{I_{7}}}=1+.. \label{s14}
           \end{eqnarray} 
           
        By adding three dimensional manifold of equation (\ref{s14}) to the eleven dimensional manifold of equation (\ref{s13}), we get:

            \begin{eqnarray}
            && \int_{M^{N=3}} \times \int_{M^{11}}\sqrt{g}\Big(G_{I_{1}I_{2}I_{3}I_{4}}G^{I_{1}I_{2}I_{3}I_{4}}\Big) = \nonumber\\&& \int_{M^{11}+y^{I_{5}}+y^{I_{6}}+y^{I_{7}}}\sqrt{g}\epsilon^{I_{4}I_{5}}\epsilon^{I_{4}I_{6}}\epsilon^{I_{5}I_{7}}\epsilon^{I_{6}I_{7}} G_{I_{1}I_{2}I_{3}I_{4}}G_{I_{1}I_{2}I_{3}I_{4}}\frac{\partial X^{I_{5}}}{\partial y^{I_{5}}}\frac{\partial X^{I_{5}}}{\partial y^{I_{6}}}\frac{\partial X^{I_{7}}}{\partial y^{I_{7}}}= \nonumber\\&&  \int_{M^{11}+M^{N=3}}\sqrt{g}CGG \nonumber\\&&\nonumber\\&&\Rightarrow C_{I_{5}I_{6}I_{7}}= \Sigma_{I_{5}I_{6}I_{7}}\epsilon^{I_{5}I_{6}I_{7}}\frac{\partial X^{I_{5}}}{\partial y^{I_{5}}}\frac{\partial X^{I_{5}}}{\partial y^{I_{6}}}\frac{\partial X^{I_{7}}}{\partial y^{I_{7}}}\label{s15}
            \end{eqnarray} 
            
        This equation has three interesting results : 1. CGG terms may be appeared in the action of supergravity by adding a three dimensional manifold, related to Lie-three-algebra to eleven dimensinal supergravity. 2. 11-dimensional manifol + three-Lie-algebra=14-dimensional supergravity. 3. The shape of C-terms is now clear in terms of string fields ($X^{i}$).
        
       To examine the correctness of theory we should re-obtain the gauge variation of the CGG-action in equation (\ref{s3}) in terms of fields strenths and curvatures . To this end, using equation (\ref{s14} and \ref{s15}), we can calculate the gauge variation of C \cite{A1}:

             \begin{eqnarray}
             && X^{I}=y^{I}+A^{I}\Rightarrow \frac{\partial \delta_{A} X^{I}}{\partial y^{I}}=\delta ( y^{I}) \nonumber\\&&\nonumber\\&&\Rightarrow \int_{M^{N=3}+M^{11}}\delta_{A} C_{I_{5}I_{6}I_{7}}=\int_{M^{N=3}+M^{11}} \Sigma_{I_{5}I_{6}I_{7}}\epsilon^{I_{5}I_{6}I_{7}}\delta_{A}(\frac{\partial X^{I_{5}}}{\partial y^{I_{5}}}\frac{\partial X^{I_{5}}}{\partial y^{I_{6}}}\frac{\partial X^{I_{7}}}{\partial y^{I_{7}}})=\nonumber\\&& \int_{M^{N=3}+M^{10}}\Sigma_{I_{5}I_{6}}\epsilon^{I_{5}I_{6}}(\frac{\partial X^{I_{5}}}{\partial y^{I_{5}}}\frac{\partial X^{I_{6}}}{\partial y^{I_{6}}})=\nonumber\\&&\int_{M^{N=3}+M^{10}}(F^{I_{5}I_{6}}-R^{I_{5}I_{6}}+\partial_{I_{5}}\phi\partial_{I_{6}}\phi)\label{s16}
             \end{eqnarray}

                  Using above equation and   equation (\ref{s13})  we can calculate the gauge variation of the CGG action in equation of (\ref{s15}):
                 
                 \begin{eqnarray}
                      &&  \delta\int_{M^{11}+M^{N=3}}\sqrt{g}CGG = \nonumber\\&& \delta\int_{M^{11}+M^{N=3}}\sqrt{g}\epsilon^{I_{1}I_{2}I_{3}I_{4}I'_{1}I'_{2}I'_{3}I'_{4}I_{5}I_{6}I_{7}}\epsilon^{I_{5}I_{6}I_{7}}(\frac{\partial X^{I_{5}}}{\partial y^{I_{5}}}\frac{\partial X^{I_{6}}}{\partial y^{I_{6}}}\frac{\partial X^{I_{7}}}{\partial y^{I_{7}}}) G_{I_{1}I_{2}I_{3}I_{4}}G_{I'_{1}I'_{2}I'_{3}I'_{4}}= \nonumber\\&&\delta\int_{M^{11}+M^{N=3}}\sqrt{g}\epsilon^{I_{1}I_{2}I_{3}I_{4}I'_{1}I'_{2}I'_{3}I'_{4}I_{5}I_{6}I_{7}}\epsilon^{I_{5}I_{6}I_{7}}(\frac{\partial X^{I_{5}}}{\partial y^{I_{5}}}\frac{\partial X^{I_{6}}}{\partial y^{I_{6}}}\frac{\partial X^{I_{7}}}{\partial y^{I_{7}}}) \times \nonumber\\&&(\epsilon^{I_{1}I_{2}I_{3}I_{4}}\frac{\partial X^{I_{1}}}{\partial y^{I_{1}}}\frac{\partial X^{I_{2}}}{\partial y^{I_{2}}}\frac{\partial X^{I_{3}}}{\partial y^{I_{3}}}\frac{\partial X^{I_{4}}}{\partial y^{I_{4}}})(\epsilon^{I'_{1}I'_{2}I'_{3}I'_{4}}\frac{\partial X^{I'_{1}}}{\partial y^{I'_{1}}}\frac{\partial X^{I'_{2}}}{\partial y^{I'_{2}}}\frac{\partial X^{I'_{3}}}{\partial y^{I'_{3}}}\frac{\partial X^{I'_{4}}}{\partial y^{I'_{4}}})=\nonumber\\&& \int_{M^{10}+M^{N=3}}\sqrt{g}\epsilon^{I_{1}I_{2}I_{3}I_{4}I'_{1}I'_{2}I'_{3}I'_{4}I_{5}I_{6}I_{7}}\epsilon^{I_{5}I_{6}}(\frac{\partial X^{I_{5}}}{\partial y^{I_{5}}}\frac{\partial X^{I_{6}}}{\partial y^{I_{6}}}) \times \nonumber\\&&(\epsilon^{I_{1}I_{2}I_{3}I_{4}}\frac{\partial X^{I_{1}}}{\partial y^{I_{1}}}\frac{\partial X^{I_{2}}}{\partial y^{I_{2}}}\frac{\partial X^{I_{3}}}{\partial y^{I_{3}}}\frac{\partial X^{I_{4}}}{\partial y^{I_{4}}})(\epsilon^{I'_{1}I'_{2}I'_{3}I'_{4}}\frac{\partial X^{I'_{1}}}{\partial y^{I'_{1}}}\frac{\partial X^{I'_{2}}}{\partial y^{I'_{2}}}\frac{\partial X^{I'_{3}}}{\partial y^{I'_{3}}}\frac{\partial X^{I'_{4}}}{\partial y^{I'_{4}}})=\nonumber\\&& \int_{M^{10}+M^{N=3}}\sqrt{g}\epsilon^{I_{1}I_{2}I_{3}I_{4}I'_{1}I'_{2}I'_{3}I'_{4}I_{5}I_{6}I_{7}}(\epsilon^{I_{4}I_{5}}\frac{\partial X^{I_{4}}}{\partial y^{I_{4}}}\frac{\partial X^{I_{5}}}{\partial y^{I_{5}}})(\epsilon^{I'_{4}I_{6}}\frac{\partial X^{I'_{4}}}{\partial y^{I_{4}}}\frac{\partial X^{I_{6}}}{\partial y^{I_{6}}}) \times \nonumber\\&&(\epsilon^{I_{1}I_{2}}\frac{\partial X^{I_{1}}}{\partial y^{I_{1}}}\frac{\partial X^{I_{2}}}{\partial y^{I_{2}}})(\epsilon^{I'_{1}I'_{2}}\frac{\partial X^{I'_{1}}}{\partial y^{I'_{1}}}\frac{\partial X^{I'_{2}}}{\partial y^{I'_{2}}})(\epsilon^{I_{3}I'_{3}}\frac{\partial X^{I_{3}}}{\partial y^{I_{3}}}\frac{\partial X^{I'_{3}}}{\partial y^{I'_{3}}})= \nonumber\\&& \int_{M^{10}+M^{N=3}}\sqrt{g}\Sigma_{n=1}^{5}\Big(tr F^{n}-tr R^{n}+trF^{n}R^{5-n}\Big) + \nonumber\\&& \delta\int_{M^{10}+M^{N=3}}\sqrt{g}\Sigma_{m=1}^{5}\Sigma_{n=1}^{5}\Big((\partial \phi)^{2m}tr R^{5-2m} +(\partial \phi)^{2m}trF^{n}R^{5-n-2m}-\frac{1}{2}\partial_{\mu}\phi\partial^{\mu}\phi\Big)+..\label{s17}
                      \end{eqnarray} 
                       
               In above equation, the first part of results cancel the anomaly in equation (\ref{s3}), however the second part produces the action in extended theories of gravity. In fact, we can rewrite the second part of above results in following form:

                         \begin{eqnarray}
                         && S=\int_{M^{11}+M^{N=3}}\sqrt{g}\Big(F(R,\phi)-\frac{1}{2}\partial_{\mu}\phi\partial^{\mu}\phi\Big)+.. \nonumber\\&& F(R,\phi)=\Sigma_{m=1}^{5}\Sigma_{n=1}^{5}\Big((\partial \phi)^{2m}tr R^{5-2m} +(\partial \phi)^{2m}trF^{n}R^{5-n-2m}\Big) \label{ps16}
                         \end{eqnarray}
                            
  This is a version of the action in extended theories of gravity \cite{kj3,kj4} which is obtained from G-theory. Thus, G-theory not only help us to remove anomalies in ten and eleven -dimensional supergravity but also yields the predicted forms of the action in Dark energy models.  
    
\section{Summary and conclusion }\label{o4}
In this research, we have investigated the origin of Padmanabhan mechanism in G-theory which is more complete respect to M-theory and reduces to it in some limitations. Until now, M-theory was known as the mother of different types of string theory. This theory is an extended version of superstring theory in eleven dimensions that by compactification can be reduced to type IIA and $E_{8}\times E_{8}$ heterotic string theory. Then these theories can be transited to other string theories by various string dualities. M-theory has some limitations e.q., some needed objects like M3-branes which our universe may be placed on it, are unstable. Also, the reason for emergence only M2 and M5 branes and the supersymmetry is not clear. For this reason, we have to construct one bigger theory which contains more stable objects without limitation of M-theory. We introduce G-theory as  a new model  which other theories like  M-theory and superstring theory are originated from it. In this model, first, two types of scalar strings, one with positive energy and one with negative energy are produced and construct new object, named G0-branes in fourteen dimensions.  Then, by compacting these branes  on three circles via two different ways (symmetrically
  and anti-symmetrically), two bosonic and fermionic parts of action for M0-branes are created and G-theory transits to M-theory. Finally, these M0-branes link to each other and produce  supersymmetric Mp-branes  which contain the same  number of degrees of freedom for both fermions and bosons. Coincidence with the birth of Mp-branes, our universe is born on one of them and interact with extra energy and another branes. Number of degrees of freedom on this universe depends on the energy of Mp-brane and number of degrees of freedom in the bulk is related to the energy of other Mp-brane and extra energy. This extra energy dissolves in our universe and leads to an increase in number of degrees of freedom on it and expansion of universe. We obtained the magnitude of the slow-roll parameters and the tensor-to-scalar ratio and found that they were very smaller than one which is in agreement with predictions of experiments. Finally, we obtained the explicit form of the actions in extended theories of gravity in G-theory and showed that these results are in agreement with results of \cite{kj3,kj4} .

\section*{Acknowledgments}
\noindent  The work of Alireza Sepehri has been supported financially by Research Institute for Astronomy-Astrophysics of Maragha (RIAAM) under research project No.1/4165-21. The work was partly supported by VEGA Grant No. 2/0009/16. R. Pincak would like to thank the TH division in CERN for hospitality.

\end{document}